\documentclass[aps,12pt,eqsecnum,preprint,nofootinbib,superscriptaddress,a4paper]{revtex4}
\usepackage{amssymb,amsmath,amsthm,graphicx,amscd}
\usepackage[mathscr]{eucal}
\usepackage{mathtools, enumerate,color,comment, cancel}
\usepackage{bm,slashed}
\usepackage[hidelinks]{hyperref}

\usepackage{orcidlink}

\usepackage{xcolor}

\DeclareMathSymbol{\shortminus}{\mathbin}{AMSa}{"39}

\newcommand{\fivedots}[1]{\stackrel{\scriptscriptstyle\boldsymbol{\cdot}\boldsymbol{\cdot}\boldsymbol{\cdot}\boldsymbol{\cdot}\boldsymbol{\cdot}}{#1}}

\begin{document}

\title{Non-Markovian Quantum Master and Fokker-Planck Equation for Gravitational Systems and Gravitational Decoherence}
\author{Hing-Tong Cho\orcidlink{0000-0002-8497-1490}}
\email{htcho@mail.tku.edu.tw}
\affiliation{Department of Physics, Tamkang University, Tamsui, New Taipei City, Taiwan, ROC}
\author{Bei-Lok Hu\orcidlink{0000-0003-2489-9914}}
\email{blhu@umd.edu}
\affiliation{Maryland Center for Fundamental Physics and Joint Quantum Institute,  University of Maryland, College Park, Maryland 20742, USA}

\date{April 16, 2025}

\begin{abstract}
A quantum master equation describing the stochastic dynamics of a quantum massive system interacting with a quantum gravitational field is useful for the investigation of quantum gravitational  and quantum informational issues such as the quantum nature of gravity, gravity-induced entanglement and gravitational decoherence.   Studies of the decoherence of quantum systems  by an electromagnetic field shows that a lower  temperature environment  is more conducive to  successful quantum information processing experiments. Likewise, the quantum nature of (perturbative) gravity is far better revealed at lower temperatures than high, minimizing the corruptive effects of thermal noise. In this work, generalizing earlier results of the Markovian ABH master equation \cite{AHGraDec,Blencowe} which is valid only for high temperatures,  we derive a non-Markovian quantum master equation for the reduced density matrix, and the associated Fokker-Planck equation for the Wigner distribution function, for the stochastic dynamics of two masses following quantum trajectories, interacting with a graviton field, including the effects of graviton noise, valid for all temperatures. We follow the influence functional approach exemplified in the derivation of the non-Markovian Hu-Paz-Zhang master equation \cite{HPZ92,HM94} for quantum Brownian motion. We find that in the low temperature limit, the off-diagonal elements of the reduced density matrix decrease in time  logarithmically for the zero temperature part and quadratically in time for the temperature-dependent part, which is distinctly different from the Markovian case.  We end with a summary of our findings and a discussion on how this problem studied here is related to the quantum stochastic equation derived in \cite{GHL06} for gravitational self force studies, and to quantum optomechanics where experimental observation of gravitational decoherence and entanglement may be implemented.

\end{abstract}

\maketitle

\tableofcontents

\section{Introduction}

A strong motivating force for our present endeavor to derive a quantum master equation for a quantum description of quantum massive systems interacting with a quantum gravitational field is for the investigation of quantum gravitational (QG) and quantum informational (QI) issues such as gravitational decoherence \cite{AHGraDec,Blencowe,Wang,LagAna} \footnote{For a thematic overview, see, e.g., \cite{AHTheme}. The theoretical foundation of our discussions here is based on general relativity (GR) and quantum field theory (QFT) solely, allowing for no alternatives.  For a topical review which covers models where quantum theory is modified, under the assumption that gravity causes the `collapse of the wave function' when systems are large enough, see  \cite{Bassi}. For earlier work on decoherence in quantum gravity, including `intrinsic or `fundamental decoherence, see \cite{Milburn,Gambini} and their critiques \cite{AHDecQG}. For decoherence and dephasing effects due to metric fluctuations,  see, e.g., \cite{JacRey,Ford,Boni,Breuer} and references cited in the sample papers mentioned above.  For a pedagogical introduction to the quantum field theory of gravitons, graviton noise and gravitational decoherence, see a recent tutorial \cite{HCH24}.}, the quantum nature of gravity in proposed laboratory experiments \cite{QGPhen,Sabine,Carney,Huggett}, in gravitational wave phenomena \cite{Pang,Guerreiro20,McCuller} and early universe cosmology \cite{GriSid,Weinberg05,Hertzberg}, the detection of gravitons \cite{Dyson}, the observational effects of graviton noise \cite{CalHu94,PWZ,Kanno,ChoHu22,ChoHu23,Suzuki} and  gravity-induced entanglement \cite{Bose,Vedral,AHComment,AHGravCat,Wald,Christ}, to name a few topics of current interest.  We shall focus on gravitational decoherence as a first application of our derived equation here,  leaving a vast number of foundational issues pertaining to QI in QG to later investigations. 

Studies of gravitational decoherence ask the question: how does a quantum gravitational system,  as simple as, say, two masses (described by wave functions or wavepackets) in some quantum states evolve to start acquiring classical attributes from their interactions  with a quantum gravitational field (or gravitons, the quantized weak linear perturbations off of a background spacetime). Decoherence has been studied and demonstrated for atomic systems interacting with an electromagnetic field \cite{Haroche,Wineland}. The gravitational force, though feeble by comparison, is ubiquitous, as  gravitational interactions accompany any massive object, even massless ones (light). From a foundational theoretical perspective, it has been suggested \cite{Penrose} that the way to resolve the incongruence between quantum and gravity is not to quantize general relativity but to `gravitise' quantum mechanics. Therewith,  we see the special role gravity plays, not only in governing the interaction of massive systems, but also in establishing the foundations of the quantum structure. (See also \cite{Adler}.)  

Decoherence in the environment-induced vein \cite{Joos,UnrZur,Zurek,Schloss} is constructed within the theoretical framework of open quantum systems \cite{OQS} \footnote{There are other pathways to study decoherence phenomena, less explicit in highlighting the features of the environments.  This includes the consistent history \cite{conhis} scheme of Griffiths and Omnes  and the decoherent history \cite{dechis} scheme of Gell-Mann and Hartle, which advocate that the notion of histories is the right way to address foundational issues of quantum mechanics and, in particular, for closed systems \cite{Har93} like the universe \cite{DecCos}. Technically, the relation between the decoherent histories formalism and the closed-time-path (CTP)  formalism has been made explicit in the correlation histories formalism of Calzetta and Hu \cite{CalHuCorHis}. For a brief description of how these different quantum decoherence formalisms are used to address quantum information issues in cosmology, see, e.g., Sec. 2 of \cite{HHDecCos} and \cite{HHEntCos}.}. Quantum master equations for the reduced density matrices describing their dynamics, together with the quantum Langevin equations  for the operators of their dynamical variables,  and the Fokker-Planck (FP) equations for their probability distribution functions,  are at its core. Quantum master equations have been studied and applied successfully to quantum optics since the 50s.   The version popularly employed there is derived under the Born (lowest perturbative order) - Markov (memoryless) approximations \cite{Walls}. There is little motivation for quantum opticians to explore non-Markovian effects because, as is often explained by them, the response of the (two-level, idealized) atom to field excitations is fast enough so they can safely ignore  memory effects. (This is not true when quantum field-induced effects need be accounted for, such as in the calculation of the entanglement between two atoms in a common bath, see, e.g., \cite{ASH06,LinHu09}.) Three other master equations of significance appeared in the following three decades: 1)  the master equation derived in 1976 by Lindblad \cite{Lindblad} and by Gorini, Kossakowski \& Sudarshan \cite{GKS},  often referred to as the Lindblad or GKS-L equation, is attractive  because of its nice mathematical properties, namely, the mapping of the open system dynamics is trace-preserving and completely positive for any initial condition.   The GKS-L equation is  popular also because it can describe a broad class of Markovian dynamics. However, the GKS-L equation has pathologies, e.g., it violates the uncertainty principle at very low temperatures \cite{LinbPatho}. 2) The Caldeira-Leggett (CL) master equation \cite{CalLeg83} derived in 1983 for the quantum Brownian motion (QBM) model which is also Markovian, valid for Ohmic environments at very high temperatures. The C-L equation has pathology as it is not a positive dynamical map \cite{HomaCL}. This is not an issue in  3) The non-Markovian Hu-Paz-Zhang (HPZ) master equation \cite{HPZ92,HPZ93} derived in 1992 (see also \cite{HM94,HalYu96})  for the QBM model, valid for a general (Ohmic and non-Ohmic) environment at all temperatures (see the analysis of \cite{Homa20,Homa23}). Environments with colored noises or with multiple time scales are nowadays often called non-Markovian environments. Newer results about decoherence at very low temperatures were obtained, e.g., in \cite{PHZ93}, with this equation.  Because it is free of the pathologies of the above mentioned two most popular Markovian equations, the HPZ equation has served as a  benchmark for the study of non-Markovian effects,  for systems describable by the generic QBM model.

We mentioned these three master equations because they are the most often invoked for open quantum system dynamics in a broad range of disciplines, and, for our specific purpose here, they together provide a useful platform to address several important issues:   1) {\it The differences between   Markovian and   non-Markovian behaviors}: By comparing e.g., the C-L equation with the HPZ equation we can clearly identify where memory effects arise and make a difference in physical reality. There,  the anomalous quantum diffusion term in the HPZ equation, which is amiss in the C-L equation, increases in importance with lowering temperature, thus enabling the system to have a longer decoherence time, meaning, to maintain its quantum nature a bit longer. While skillfully making use of entanglement as a resource is at the heart of all quantum information processing schemes, keeping the system of interest quantum is a necessary condition and  maximizing its quantum coherence time is a crucial factor to consider in their experimental designs.  
2) How {\it gravitational decoherence happening in the energy basis} differs from the more commonly encountered decoherence which happens in the configuration space. Many familiar examples of the latter case invoke the above-mentioned master equations, such as the HPZ equation used in a generic example shown in \cite{HPZ92,PHZ93}, namely,   how a density matrix function initially with two pronounced distributions at $x= \pm L$ evolves into one largely centered at $x=0$, shredding off its off-diagonal components and settling down into a near-diagonal reduced density matrix in a finite, often very short time which we can use to define a decoherence time. Notice that the decoherence times vary greatly with the type of interaction between the system and its environment (see the plots of the Wigner function in \cite{PHZ93}). For gravitational decoherence, the quantum master equation, known as the ABH master equation,  derived by Anastopoulos \& Hu (AH) and Blencowe (B), is Markovian. These authors show that gravitational decoherence proceeds in the energy basis. The two groups differ in their estimates of the decoherence time scale under typical conditions: Blencowe's estimate is much more optimistic than that of AH, who also presented two scenarios, one pertaining to gravity as a fundamental theory and the other if gravity is emergent.  3)   Note that the ABH equation is in the Lindblad form, thus one can borrow many known properties of the Lindblad equation for its analysis, (but at the same time,  be mindful of its shortcomings at very low temperatures). The ABH equation being Markovian is valid only at sufficiently high temperatures\footnote{Note that while Blencowe assumed the Born-Markov approximation from the beginning, the derivation of Anastopoulos and Hu is valid for all temperatures (see their Eq. (52), which contains the coth factor signifier). Only in the last steps when AH took the high temperature limit did  the master equation turn Markovian.  It is this last form, in combination with the findings of Blencowe, which was referred to as the ABH equation.}. The low temperature regime where non-Markovian behaviors become important is what this paper aims to supplant.  4) The  method used in the derivation of the  HPZ equation for the study of non-Markovian environment and processes is followed here \cite{HM94} in our derivation of a non-Markovian quantum master equation for gravitational systems. 

These comparisons and contrasts set the stage for our present investigation. For quantum gravitational systems at least two issues need be addressed: a) How is gravitational decoherence different from the decoherence processes most often studied using the harmonic oscillator QBM model or models of two-level atoms interacting with an EM field?   b) How would the  non-Markovian behaviors in this  more general master equation differ  from the known behavior based on the known Markovian (ABH) master equations?   Gravitational decoherence happens in the energy basis, not in the configuration space basis, as  in common examples,  because all gravitational couplings are through the mass-energy.  We can see this from the interaction Lagrangian where the stress-energy tensor $T_{\mu\nu}$ of the system (e.g. comprised of two masses) is coupled to a weak gravitational (or graviton) field $h_{\mu\nu}$. Thus the central issue we need to address and the main task of this work is non-Markovianity,  in both the environment, namely, colored noises, and in the open system dynamics, namely, nonlocal dissipation, both exhibiting the presence of, even the prominence, of memory.  

To go beyond the Markovian limitations, we ask, in what parameter regimes would non-Markovian effects become important?  We shall present a detailed analysis later but here is a simplified answer:  Since a quantum field acting as the environment has Ohmic spectral density (see, e.g., \cite{UnrZur,HM94}), the nM behavior of special interest to us in this work manifests in the low temperature regime. As mentioned earlier, this regime is most conducive to successful quantum information processing experiments, and, here, from a theoretical viewpoint, revealing the quantum nature of (perturbative) gravity and for studying graviton processes.  We shall  endeavor to obtain a non-Markovian master equation valid for the full temperature range.

The technical tools we employ here are from the Feynman-Vernon influence functional formalism \cite{FeyVer63,CalLeg83,HPZ92} or the formally related Schwinger-Keldysh \cite{Sch61,Kel64} closed-time-path (CTP) integral or `in-in' formalism \cite{Chou85,Jor86,CalHu87,CalHu88,Weinberg05}.   The central characters are the noise and dissipation kernels in the influence action which are in  general  nonlocal, namely, colored noises in the environment and nonlocal dissipation in the open system dynamics.  To see the Markovian limit we need to sort out the relative weights of a number of competing factors, namely, the two-time difference, the inverse cutoff frequency and the inverse temperature.  This will be shown in Sec.~II after a detailed analysis of these two kernels, in particular, the noise kernel (the dissipation kernel is independent of temperature). Since the noise kernel governing quantum diffusion is responsible for decoherence, we can answer one of the key questions raised earlier, i.e., what is the behavior of decoherence at very low temperatures, or how do non-Markovian effects enter in gravitational decoherence. 

For a classical system we can readily obtain a Langevin equation by taking the functional  variation of the influence action.  It is of a semiclassical nature in that the dynamics of the system, here represented by two classical masses, is affected by the graviton noises of the quantum field environment.  We show this derivation in Sec.~IIB.  

In Sec.~III We proceed to derive a non-Markovian quantum master equation for the dynamics of a quantum open system, here also made up of two masses, but following quantum trajectories, interacting with a graviton field. The consideration of quantum masses is needed for addressing quantum information issues related to gravity such as gravitational decoherence and gravity-induced quantum entanglement. As has become known from other well-studied systems, such as for the generic QBM model, the derivation of a non-Markovian quantum master equation is considerably more involved than of the Markovian master equation.  We follow the route mapped out in Hu and Matacz \cite{HM94} where details can be found in the Appendices there.   In Sec.~III we derive the evolution  operator of the reduced density matrix of a quantum two-mass system (reduced to one mass by adopting a Fermi normal coordinate along one of the masses). In Sec.~IV we derive the quantum master equation for this system and highlight its non-Markovian features. We then derive the decoherence time in the non-Markovian regime and compare with the results obtained earlier in the Markovian regime. In Sec~V  we summarize the settings and findings of this paper and mention three topics which the present work are connected to, as possible future  developments.  Some technical details are relegated to the Appendices.


\section{Influence functional of a quantum gravitational field on the motion of two masses}\label{S:eiwt}
In this section we briefly describe the derivation of the influence functional due to the graviton bath following Refs.~\cite{ChoHu22,ChoHu23}. More background material on the quantum field theory of gravitons and graviton noise can be found in \cite{HCH24} and references cited therein. To begin with,  we consider the Einstein action
\begin{eqnarray}
S_{g}=\frac{1}{\kappa^{2}}\int d^{4}x \sqrt{-g}\,R
\end{eqnarray}
where $\kappa^{2}=16\pi G$ and $G$ is the Newton's constant. Gravitons are quantized linear perturbations of a background spacetime, here assumed to be the Minkowski space. See Refs.~\cite{GalHu05,GHL06} for the influence functional due to a scalar field and a vector field in a curved background spacetime.  It has been shown, for example, in \cite{ForPar77}, that the dynamics of gravitons can be represented by two minimally coupled massless scalar fields $h^{(s)}(x)$, where $s$ denotes the two graviton polarizations. The corresponding action can be expressed as 
\begin{eqnarray}\label{gaction}
S_{grav}=-\frac{1}{2}\int\,d^{4}x\,\sum_{s}\partial_{\alpha}h^{(s)}(x)\partial^{\alpha}h^{(s)}(x)
\end{eqnarray}

Since we are interested in the effects of a quantum gravitational field on the geodesic deviation between two moving massive particles, it is natural to use the Fermi normal coordinates $(t,\vec{z})$ for these two particles. The coordinate is set up in a way where one of the two particles' geodesics, call it the reference particle,  is taken to be the origin of the Fermi normal coordinate and its temporal component to form the time axis. As such, the separation of the geodesic  of the system particle from the origin will produce the geodesic deviation. 
Expanding up to quadratic order in $\vec{z}$, we have the action for a particle with mass $m$
\begin{eqnarray}\label{maction}
S_{m}&=&-m\int\,\sqrt{-ds^{2}}\nonumber\\
&=&\int\,dt\,\left[\frac{m}{2}\,\delta_{ij}\,\dot{ z}^{i}\dot{ z}^{j}+\frac{m\kappa}{4}\,\ddot{h}_{ij}\, z^{i} z^{j}\right]+\cdots
\end{eqnarray}
The second term represents the interaction between the particle and the graviton field. 

In an open quantum systems approach \cite{OQS} based on the Feynman-Vernon \cite{FeyVer63} influence functional formalism  one integrates over the graviton field to produce the influence action, which encapsulates the effects of the graviton field on the particle motion. Since the interaction term in Eq.~(\ref{maction}) is linear in the graviton field, this influence action can be evaluated exactly giving
\begin{eqnarray}\label{IFaction2}
S_{IF}&=&\int_{0}^{t} dt_{1}\int_{0}^{t}\,dt_{2}\,\left(\frac{d^{2}}{dt_{1}^{2}}\Delta^{ij}(t_{1})\right)D_{ijkl}(t_{1},t_{2})\left(\frac{d^{2}}{dt_{2}^{2}}\Sigma^{kl}(t_{2})\right)\nonumber\\
&&\ \ +\frac{i}{2}\int_{0}^{t} dt_{1}\int_{0}^{t}\,dt_{2}\,\left(\frac{d^{2}}{dt_{1}^{2}}\Delta^{ij}(t_{1})\right)N_{ijkl}(t_{1},t_{2})\left(\frac{d^{2}}{dt_{2}^{2}}\Delta^{kl}(t_{2})\right)
\end{eqnarray}
where 
\begin{eqnarray}
\Sigma^{ij}(t)&=&\frac{1}{2}\left[ z_{+}^{i}(t) z_{+}^{j}(t)+ z_{-}^{i}(t) z_{-}^{j}(t)\right]\\\Delta^{ij}(t)&=& z_{+}^{i}(t) z_{+}^{j}(t)- z_{-}^{i}(t) z_{-}^{j}(t),
\end{eqnarray}
$ z_{+}^{i}(t)$ and $ z_{-}^{i}(t)$ are the forward and backward in-time fields in the closed-time path formalism. Moreover, with the definition of the coupling constant $\alpha=m\kappa/(2\sqrt{2}\,(2\pi)^3)$,
\begin{eqnarray}\label{dissipation}
D_{ijkl}(t,t')=\alpha^{2}\int d^{3}k\,d^{3}k'\int d^{3}x\,d^{3}x'\,e^{-i\vec{k}\cdot\vec{x}}e^{-i\vec{k}'\cdot\vec{x}'}\sum_{s}\epsilon_{ij}^{(s)}(\vec{k})\epsilon_{kl}^{(s)}(\vec{k}')G_{ret}(x,x')
\end{eqnarray}
is the dissipation kernel with $\epsilon_{ij}^{(s)}(\vec{k})$ being the graviton polarization tensor, and $G_{ret}(x,x')=i\theta(t-t')\langle[h(x),h(x')]\rangle$ being the retarded Green function of a massless scalar field $h(x)$. Also,
\begin{eqnarray}\label{noise}
N_{ijkl}(t,t')=\frac{\alpha^{2}}{2}\int d^{3}k\,d^{3}k'\int d^{3}x\,d^{3}x'\,e^{-i\vec{k}\cdot\vec{x}}e^{-i\vec{k}'\cdot\vec{x}'}\sum_{s}\epsilon_{ij}^{(s)}(\vec{k})\epsilon_{kl}^{(s)}(\vec{k}')G^{(1)}(x,x')
\end{eqnarray}
is the noise kernel with $G^{(1)}(x,x')=\langle\{h(x),h(x')\}\rangle$ being the Hadamard function. Combining the influence action $S_{IF}$ with the noninteracting part of the matter action
\begin{eqnarray}
    S_{m0}=\int\,dt\,\frac{m}{2}\,\delta_{ij}\,\dot{ z}^{i}\dot{ z}^{j}.
\end{eqnarray}
one obtains the coarse-grained effective action
\begin{eqnarray}\label{SCG}
    S_{CG}=S_{m0}[z_{+}]-S_{m0}[z_{-}]+S_{IF}
\end{eqnarray}
Note that, as compared to our previous papers \cite{ChoHu22,ChoHu23}, we have defined the dissipation and the noise kernels without the time derivatives which are left acting on $\Sigma^{ij}$ and $\Delta^{ij}$. This will be found convenient when we work on the master equation when the particles are considered to be quantum.

Using the Feynman-Vernon Gaussian functional identity, the noise part of the influence action in Eq.~(\ref{IFaction2}) can be replaced by a path integral over a stochastic tensor force $\xi_{ij}(t)$.
\begin{eqnarray}
    e^{-\frac{1}{2}\int_{0}^{t} dt_{1}\int_{0}^{t}\,dt_{2}\,\left(\frac{d^{2}}{dt_{1}^{2}}\Delta^{ij}(t_{1})\right)N_{ijkl}(t_{1},t_{2})\left(\frac{d^{2}}{dt_{2}^{2}}\Delta^{kl}(t_{2})\right)}={\cal N}\int\,D\xi\,P[\xi]\,e^{i\int_{0}^{t}dt_{1}\left(\frac{d^{2}}{dt_{1}^{2}}\Delta^{ij}(t_{1})\right)\xi_{ij}(t_{1})}
\end{eqnarray}
where ${\cal N}$ is a normalization constant, and the gaussian probability density
\begin{eqnarray}
    P[\xi]={\cal N}\,e^{-\frac{1}{2}\int_{0}^{t}dt_{1}\int_{0}^{t}dt_{2}\,\xi^{ij}(t_{1})\,(N^{-1})_{ijkl}(t_{1},t_{2})\,\xi^{kl}(t_{2})}
\end{eqnarray}
The correlation function of the stochastic tensor force is therefore given by the noise kernel.
\begin{eqnarray}
    \langle\xi_{ij}(t_{1})\xi_{kl}(t_{2})\rangle_{s}&=&{\cal N}\int\,D\xi\,P[\xi]\,\xi_{ij}(t_{1})\xi_{kl}(t_{2})\nonumber\\
    &=&N_{ijkl}(t_{1},t_{2})
\end{eqnarray}
where $\langle\rangle_{s}$ represents the stochastic average. Together with the kinetic terms of the mass particle, we obtain the stochastic effective action
\begin{eqnarray}\label{seaaction}
    S_{SEA}&=&S_{m}[z_{+}]-S_{m}[z_{-}]+S_{IF}\nonumber\\
    &=&\frac{m}{2}\int dt\,\delta_{ij}\dot{ z}_{+}^{i}(t)\dot{ z}_{+}^{j}(t)-\frac{m}{2}\int dt\,\delta_{ij}\dot{ z}_{-}^{i}(t)\dot{ z}_{-}^{j}(t)\nonumber\\
&&\hskip 10pt + \int dt\,dt'\,\left(\frac{d^{2}}{dt_{1}^{2}}\Delta^{ij}(t)\right)D_{ijkl}(t,t')\left(\frac{d^{2}}{dt_{2}^{2}}\Sigma^{kl}(t')\right)+\int dt\,\xi_{ij}(t)\left(\frac{d^{2}}{dt_{1}^{2}}\Delta^{ij}(t)\right)\nonumber\\
\end{eqnarray}
In the following we shall analyze the noise and dissipation kernels and, upon taking the variations on this stochastic effective action, derive the Langevin equation of motion.

\subsection{Noise and dissipation kernels: Markovian and non-Markovian regimes}

Let us examine more closely the noise and dissipation kernels in Eqs.~(\ref{noise}) and (\ref{dissipation}), respectively, to understand their physical properties, and, in particular, to identify the Markovian regime from the non-Markovian regimes, which are the focus of our present work.   In \cite{ChoHu22}, we have considered the noise kernel corresponding to various quantum states of the graviton. Here, we concentrate on the thermal state. We shall obtain expressions for the high and low temperature (T) limits, the high T limit  falls in the Markovian regime which previous results pertain, whereas the low T limit, where non-Markovian effects are expected to be prominent,  is of special interest here.  For this, we can write
\begin{eqnarray}
    N_{ijkl}(t,t')=N_{ijkl}^{(0)}(t,t')+N_{ijkl}^{(\beta)}(t,t').
\end{eqnarray}
The first term is the Minkowski vacuum result with temperature $T=0$, while the second one is the thermal part with $\beta=1/T$. For the Minkowski vacuum, the Hadamard function is given by
\begin{eqnarray}\label{hadamard}
    G^{(1)}_{0}(x,x')=\int\,d^{3}q\,[u_{\vec{q}}(x)u^{*}_{\vec{q}}(x')+u^{*}_{\vec{q}}(x)u_{\vec{q}}(x')],
\end{eqnarray}
where the mode function $u_{\vec{q}}(x)=e^{-iqt}e^{i\vec{q}\cdot\vec{x}}/(2\pi)^{3/2}\sqrt{2k}$. Integrating over $\vec{x}$, $\vec{x}'$, $\vec{k}$, and $\vec{k}'$, 
\begin{eqnarray}
    N_{ijkl}^{(0)}(t,t')=4\pi^{3}\alpha^{2}\int\,d^{3}q\left(\frac{1}{q}\right)\cos[q(t-t')]\sum_{s}\epsilon_{ij}^{(s)}(\vec{q})\epsilon_{kl}^{(s)*}(\vec{q}).
\end{eqnarray}
Next, integrating over the solid angle $\Omega_{q}$, one obtains \cite{ChoHu22}
\begin{eqnarray}\label{sumpol}
    \int\,d\Omega_{q}\sum_{s}\epsilon_{ij}^{(s)}(\vec{q})\epsilon_{kl}^{(s)*}(\vec{q})=-\left(\frac{8\pi}{15}\right)[2\delta_{ij}\delta_{kl}-3(\delta_{ik}\delta_{jl}+\delta_{il}\delta_{jk})],
\end{eqnarray}
and 
\begin{eqnarray}
    N^{(0)}_{ijkl}(t,t')=-\frac{32\pi^4\alpha^2}{15}[2\delta_{ij}\delta_{kl}-3(\delta_{ik}\delta_{jl}+\delta_{il}\delta_{jk})]\int_{0}^{\infty}dq\,q\,\cos[q(t-t')]
\end{eqnarray}
The $q$-integral is divergent. To regularize it, we put in an ultraviolet cutoff $\Lambda$,
\begin{eqnarray}\label{noiseMin}
  N^{(0)}_{ijkl}(t,t')&=&-\frac{32\pi^4\alpha^2}{15}[2\delta_{ij}\delta_{kl}-3(\delta_{ik}\delta_{jl}+\delta_{il}\delta_{jk})]\int_{0}^{\Lambda}dq\,q\,\cos[q(t-t')]\nonumber\\
  &=&-\frac{32\pi^4\alpha^2}{15}[2\delta_{ij}\delta_{kl}-3(\delta_{ik}\delta_{jl}+\delta_{il}\delta_{jk})]\times\nonumber\\
  &&\hskip 40pt\frac{1}{(t-t')^{2}}\left\{-1+\cos[\Lambda(t-t')]+\Lambda(t-t')\sin[\Lambda(t-t')]\right\}
\end{eqnarray}

Similar consideration can be applied to the thermal part of the noise kernel \cite{ChoHu22} giving
\begin{eqnarray}\label{thermalnoise}
    &&N^{(\beta)}_{ijkl}(t,t')\nonumber\\
    &=&-\frac{64\pi^4\alpha^2}{15}[2\delta_{ij}\delta_{kl}-3(\delta_{ik}\delta_{jl}+\delta_{il}\delta_{jk})]\int_{0}^{\infty}dq\,q\,\frac{\cos[q(t-t')]}{e^{q\beta}-1}\nonumber\\
    &=&-\frac{32\pi^4\alpha^2}{15}[2\delta_{ij}\delta_{kl}-3(\delta_{ik}\delta_{jl}+\delta_{il}\delta_{jk})]\frac{1}{(t-t')^{2}}\left[1-\frac{\pi^2(t-t')^2}{\beta^2}\text{csch}^2\left(\frac{\pi(t-t')}{\beta}\right)\right]\nonumber\\
\end{eqnarray}
Note that due to the $e^{q\beta}$ factor, the $q$-integral is in fact convergent. For the low temperature expansion, we take $\pi(t-t')/\beta\ll 1$, and 
\begin{eqnarray}\label{lowexpand}
    1-\frac{\pi^2(t-t')^2}{\beta^2}\text{csch}^2\left(\frac{\pi(t-t')}{\beta}\right)
    =\frac{\pi^2(t-t')^2}{3\beta^2}-\frac{\pi^4 (t-t')^4}{15\beta^4}+\frac{2\pi^6 (t-t')^6}{189\beta^6}+\cdots\nonumber\\
\end{eqnarray}
Combining with the Minkowski part of the noise kernel, we have, in the low temperature limit,
\begin{eqnarray}\label{lowTnoise}
    N_{ijkl}(t,t')
    &=&-\frac{32\pi^4\alpha^2}{15}[2\delta_{ij}\delta_{kl}-3(\delta_{ik}\delta_{jl}+\delta_{il}\delta_{jk})]\frac{1}{(t-t')^{2}}\times\nonumber\\
  &&\hskip 20pt\bigg\{-1+\cos[\Lambda(t-t')]+\Lambda(t-t')\sin[\Lambda(t-t')]\nonumber\\
  &&\hskip 40pt +\frac{\pi^2(t-t')^2}{3\beta^2}-\frac{\pi^4 (t-t')^4}{15\beta^4}+\frac{2\pi^6 (t-t')^6}{189\beta^6}+\cdots\bigg\}
\end{eqnarray}
In fact, there are three scales in this expansion, $\Lambda$, $(t-t')$, and $\beta$. In making the low temperature expansion, we have assumed that $1/(t-t')\gg 1/\beta$. For the cutoff $\Lambda$, it is also understood that $\Lambda\gg 1/(t-t')$. Hence, we have for the expansion above $\Lambda\gg 1/(t-t')\gg 1/\beta$.

Regarding the three scales mentioned above, there will be a subtlety involving the high temperature expansion of the noise kernel. If we expand the result in Eq.~(\ref{thermalnoise}) for $\pi(t-t')/\beta\gg 1$, we have
\begin{eqnarray}\label{thermalint}
    &&\frac{1}{(t-t')^{2}}\left[1-\frac{\pi^2(t-t')^2}{\beta^2}\text{csch}^2\left(\frac{\pi(t-t')}{\beta}\right)\right]\nonumber\\
    &=&\frac{1}{(t-t')^{2}}-\frac{4\pi^2}{\beta^2}\left(e^{-2\pi(t-t')/\beta}+2\,e^{-4\pi(t-t')/\beta}+3\,e^{-6\pi(t-t')/\beta}+\cdots\right)
\end{eqnarray}
Note that basically the ultraviolet cutoff in the $q$-integral in Eq.~(\ref{thermalnoise}) has been taken to infinity. Therefore, we have implicitly $\Lambda\gg 1/\beta\gg 1/(t-t')$ in this expansion. Now, if we take the temperature to infinity or $\beta\rightarrow 0$ in Eq.~(\ref{thermalint}), we obtain $\pi^2/3\beta^2$ when $t=t'$ and $1/(t-t')^2$ when $t\neq t'$. Hence, we see that with $\Lambda\gg 1/\beta\gg 1/(t-t')$, the high temperature limit of the thermal part of the noise kernel here is not proportional to a delta function or its derivatives, and the limit is non-Markovian. Finally, with the Minkowski part of the noise kernel, the high temperature expansion in this case can be expressed as
\begin{eqnarray}\label{high1}
    &&N_{ijkl}(t,t')\nonumber\\
    &=&-\frac{32\pi^4\alpha^2}{15}[2\delta_{ij}\delta_{kl}-3(\delta_{ik}\delta_{jl}+\delta_{il}\delta_{jk})]\frac{1}{(t-t')^{2}}\bigg\{\cos[\Lambda(t-t')]+\Lambda(t-t')\sin[\Lambda(t-t')]\nonumber\\
    &&\hskip 30pt -\frac{4\pi^2(t-t')^2}{\beta^2}\left(e^{-2\pi(t-t')/\beta}+2\,e^{-4\pi(t-t')/\beta}+3\,e^{-6\pi(t-t')/\beta}+\cdots\right)\bigg\}
\end{eqnarray}

Obviously, we could have another high temperature expansion in which $1/\beta\gg\Lambda\gg 1/(t-t')$ with the temperature as the largest scale. In this case, we need to reexamine the integral in Eq.~(\ref{thermalnoise}). Putting in an ultraviolet cutoff $\Lambda$ in the $q$-integral and with $1\gg\beta\Lambda$, $q\beta$ is always small and we can expand the exponential in the denominator giving
\begin{eqnarray}\label{high2}
    \int_{0}^{\Lambda}dq\,q\,\frac{\cos[q(t-t')]}{e^{q\beta}-1}&=&\int_{0}^{\Lambda}dq\,\cos[q(t-t')]\left(\frac{1}{\beta}-\frac{q}{2}+\frac{q^2\beta}{12}
    +\cdots\right)
\end{eqnarray}
Consider the first term in the above expansion.
\begin{eqnarray}\label{regdelta}
    \int_{0}^{\Lambda}dq\cos[q(t-t')]=\frac{\sin[\Lambda(t-t')]}{(t-t')}\equiv\pi\delta_{\Lambda}(t-t')
\end{eqnarray}
where we have defined the regularized delta function $\delta_{\Lambda}(t-t')$ which approaches the Dirac delta function $\delta(t-t')$ as $\Lambda\rightarrow\infty$. The integral with the ultraviolet cutoff $\Lambda$ can thus be viewed as a regularized representation of the Dirac delta function. As we shall use this representation in our subsequent analysis in this paper, we shall elaborate more on it in Appendix A. Continuing with the expansion in Eq.~(\ref{high2}), we see that the second term exactly cancels with the Minkowski part of the noise kernel in Eq.~(\ref{noiseMin}). For the third term, using the regularization as in Eq.~(\ref{regdelta}) (see also Appendix A),
\begin{eqnarray}
     \int_{0}^{\Lambda}dq\,q^2\cos[q(t-t')]&=&-\frac{d^{2}}{dt^{2}}\int_{0}^{\Lambda}dq\,\cos[q(t-t')]\nonumber\\
     &=&-\pi\,\ddot{\delta}_{\Lambda}(t-t')
\end{eqnarray}
In conclusion, the high temperature limit in which $1/\beta\gg\Lambda\gg 1/(t-t')$, the noise kernel can be expanded as
\begin{eqnarray}
    N_{ijkl}(t,t')=-\frac{64\pi^5\alpha^2}{15}[2\delta_{ij}\delta_{kl}-3(\delta_{ik}\delta_{jl}+\delta_{il}\delta_{jk})]\left[\frac{1}{\beta}\delta_{\Lambda}(t-t')-\frac{\beta}{12}\ddot{\delta}_{\Lambda}(t-t')+\cdots\right].\nonumber\\
\end{eqnarray}
As $\Lambda\rightarrow\infty$, we obtain a noise kernel given by $\delta(t-t')$ and its derivatives, corresponding to a Markovian limit.

For the dissipation kernel in Eq.~(\ref{dissipation}), it involves the retarded Green function given in terms of the commutator of a scalar field. Because of this, the dissipation kernel is independent of temperature. Hence, using the mode function in Eq.~(\ref{hadamard}) and the result of the sum over the polarization tensors in Eq.~(\ref{sumpol}),
\begin{eqnarray}
    D_{ijkl}(t,t')=-\frac{64\pi^4\alpha^2}{15}[2\delta_{ij}\delta_{kl}-3(\delta_{ik}\delta_{jl}+\delta_{il}\delta_{jk})]\,\theta(t_{1}-t_{2})\int_{0}^{\infty}dq\,q\,\sin[q(t-t')]\nonumber\\
\end{eqnarray}
The integral is divergent and needs to be regularized as for the noise kernel. 
\begin{eqnarray}\label{disfinal}
    D_{ijkl}(t,t')
    &=&-\frac{64\pi^4\alpha^2}{15}[2\delta_{ij}\delta_{kl}-3(\delta_{ik}\delta_{jl}+\delta_{il}\delta_{jk})]\,\theta(t_{1}-t_{2})\left(-\frac{d}{dt}\right)\int_{0}^{\Lambda}dq\,\cos[q(t-t')]\nonumber\\
    &=&\frac{64\pi^5\alpha^2}{15}[2\delta_{ij}\delta_{kl}-3(\delta_{ik}\delta_{jl}+\delta_{il}\delta_{jk})]\,\theta(t_{1}-t_{2})\dot{\delta}_{\Lambda}(t-t')
\end{eqnarray}
given by the derivative of the regularized delta function. As $\Lambda\rightarrow\infty$, we reach the Markovian limit in which the dissipation is expressed as the derivative of a delta function.

\subsection{Langevin equation of classical mass motion in a graviton field}

The equation of motion derived from the stochastic effective action in Eq.~(\ref{seaaction}) is in the form of a Langevin equation. More specifically, we take the variation of $z_{+}^{i}$,
\begin{eqnarray}\label{varyS}
    \left.\frac{\delta S_{SEA}}{\delta z^{i}_{+}}\right|_{z_{+}=z_{-}=z}=0
\end{eqnarray}
Consider first the variation of the kinetic term of $z_{+}^{i}$ in $S_{SEA}$, 
\begin{eqnarray}
    \delta\left[\frac{m}{2}\int_{0}^{t}dt_{1}\,\delta_{ij}\,\dot{z}_{+}^{i}(t_{1})\dot{z}_{+}^{j}(t_{1})\right]=-m\int_{0}^{t}dt_{1}\delta_{ij}\left(\frac{d^{2}}{dt_{1}^{2}}z_{+}^{j}(t_{1})\right)\delta z_{+}^{i}(t_{1})
\end{eqnarray}
where we have assumed that $\delta z_{+}^{i}(0)=\delta z_{+}^{i}(t)=0$. Similarly, for the stochastic force term
\begin{eqnarray}
    \delta\left[\int_{0}^{t}dt_{1}\,\xi_{ij}(t_{1})\left(\frac{d^{2}}{dt_{1}^{2}}\Delta_{ij}\right)\right]=2\int_{0}^{t}dt_{1}\,\ddot{\xi}_{ij}(t_{1})\,z_{+}^{j}(t_{1})\,\delta z_{+}^{i}(t_{1})
\end{eqnarray}
where in addition we have taken $d(\delta z_{+}^{i}(t_{1}))/dt_{1}|_{t_{1}=0}=d(\delta z_{+}^{i}(t_{1}))/dt_{1}|_{t_{1}=t}=0$.
Lastly, when we take the variation of $z_{+}^{i}$ on the term involving the dissipation kernel, we have
\begin{eqnarray}
    &&\delta\left[\int_{0}^{t} dt_{1}\int_{0}^{t}\,dt_{2}\,\left(\frac{d^{2}}{dt_{1}^{2}}\Delta^{ij}(t_{1})\right)D_{ijkl}(t_{1},t_{2})\left(\frac{d^{2}}{dt_{2}^{2}}\Sigma^{kl}(t_{2})\right)\right]\nonumber\\
    &=&\int_{0}^{t}dt_{1}\int_{0}^{t}dt_{2}\,\Bigg\{\delta z_{+}^{i}(t_{1})\,z_{+}^{j}(t_{1})\left(\frac{d^{2}}{dt_{1}^{2}}D_{ijkl}(t_{1},t_{2})\right)\left[\frac{d^{2}}{dt_{2}^{2}}\left(z_{+}^{k}(t_{2})z_{+}^{l}(t_{2})+z_{-}^{k}(t_{2})z_{-}^{l}(t_{2})\right)\right]\nonumber\\
    &&\hskip 60pt +\left[\frac{d^{2}}{dt_{1}^{2}}\left(z_{+}^{i}(t_{1})z_{+}^{j}(t_{1})-z_{-}^{i}(t_{1})z_{-}^{j}(t_{1})\right)\right]\delta z_{+}^{k}(t_{2})\,z_{+}^{l}(t_{2})\left(\frac{d^{2}}{dt_{2}^{2}}D_{ijkl}(t_{1},t_{2})\right)\Bigg\}\nonumber\\
\end{eqnarray}
Note that when we take the limit $z_{+}^{i},z_{-}^{i}\rightarrow z^{i}$, the second term vanishes and it will not contribute to the equation of motion.

Putting the result of the variations of all the terms in the stochastic effective action, we obtain the equation of motion for $z_{i}$ as indicated in Eq.~(\ref{varyS}),
\begin{eqnarray}\label{langevin}
    m\delta_{ij}\,\ddot{z}^{j}(t_{1})-2\,z^{j}(t_{1})\,\frac{d^{2}}{dt_{1}^{2}}\int_{0}^{t}dt_{2}\,D_{ijkl}(t_{1},t_{2})\,\frac{d^{2}}{dt_{2}^{2}}(z^{k}(t_{2})z^{l}(t_{2}))=2\ddot{\xi}_{ij}(t_{1})\,z^{j}(t_{1})
\end{eqnarray}
which is indeed in the form of a Langevin equation with the stochastic tensor force $\xi_{ij}$. The term with the dissipation kernel involves higher derivatives as in a Mino-Sasaki-Tanaka-Quinn-Wald (MST-QW) like equation for gravitational self force \cite{MST,QW}. Moreover, in the case of a graviton field the dissipation kernel is given by Eq.~(\ref{disfinal}) as the time derivative of a delta function, that is, 
\begin{eqnarray}\label{intdis}
    &&\int_{0}^{t}dt_{2}\,D_{ijkl}(t_{1},t_{2})\,\frac{d^{2}}{dt_{2}^{2}}(z^{k}(t_{2})z^{l}(t_{2}))\nonumber\\
    &=&\frac{64\pi^4\alpha^2}{15}[2\delta_{ij}\delta_{kl}-3(\delta_{ik}\delta_{jl}+\delta_{il}\delta_{jk})]\int_{0}^{t_{1}}dt_{2}\,\dot{\delta}(t_{1}-t_{2})\frac{d^{2}}{dt_{2}^{2}}(z^{k}(t_{2})z^{l}(t_{2}))
\end{eqnarray}
Applying the result for the integral $I_{1}$ in Eq.~(\ref{intI1}) to the $t_{2}$-integral, we have
\begin{eqnarray}
    &&
    \int_{0}^{t_{1}}dt_{2}\,\dot{\delta}(t_{1}-t_{2})\frac{d^{2}}{dt_{2}^{2}}(z^{k}(t_{2})z^{l}(t_{2}))\nonumber\\
    &=&\delta(0)\frac{d^{2}}{dt_{1}^{2}}(z^{k}(t_{1})z^{l}(t_{1}))-\delta(t_{1})\left[\frac{d^{2}}{dt_{1}^{2}}(z^{k}(t_{1})z^{l}(t_{1})\right]_{t_{1}\rightarrow 0}\nonumber\\
    &&\hskip 60pt 
    -\frac{1}{2}\left[\theta(t_{1})-\theta(-t_{1})\right]\frac{d^{3}}{dt_{1}^{3}}(z^{k}(t_{1})z^{l}(t_{1}))
\end{eqnarray}
and 
\begin{eqnarray}
    &&z^{j}(t_{1})\frac{d^{2}}{dt_{1}^{2}}\left[\int_{0}^{t_{1}}dt_{2}\,\dot{\delta}(t_{1}-t_{2})\frac{d^{2}}{dt_{2}^{2}}(z^{k}(t_{2})z^{l}(t_{2}))\right]\nonumber\\
    &=&\delta(0)\,z^{j}(t_{1})\frac{d^{4}}{dt_{1}^{4}}(z^{k}(t_{1})z^{l}(t_{1}))-3\,\delta(t_{1})\left[z^{j}(t_{1})\frac{d^{4}}{dt_{1}^{4}}(z^{k}(t_{1})z^{l}(t_{1}))\right]_{t_{1}=0}\nonumber\\
    &&\ \ -3\left(\frac{d}{dt_{1}}\delta(t_{1})\right)\left[z^{j}(t_{1})\frac{d^{3}}{dt_{1}^{3}}(z^{k}(t_{1})z^{l}(t_{1}))\right]_{t_{1}=0}\nonumber\\
    &&\ \ -\left(\frac{d^{2}}{dt_{1}^{2}}\delta(t_{1})\right)\left[z^{j}(t_{1})\frac{d^{2}}{dt_{1}^{2}}(z^{k}(t_{1})z^{l}(t_{1}))\right]_{t_{1}=0}\nonumber\\
    &&\ \ -\frac{1}{2}[\theta(t_{1})-\theta(-t_{1})]\,z^{j}(t_{1})\frac{d^{5}}{dt_{1}^{5}}(z^{k}(t_{1})z^{l}(t_{1}))
\end{eqnarray}
With the tensor structure of $D_{ijkl}(t_{1},t_{2})$ as shown in Eq.~(\ref{intdis}), 
\begin{eqnarray}
    z^{k}(t_{1})z^{l}(t_{1})[2\delta_{ij}\delta_{kl}-3(\delta_{ik}\delta_{jl}+\delta_{il}\delta_{jk})]&=&-6\left(z^{i}(t_{1})z^{j}(t_{1})-\frac{1}{3}\delta^{ij}\delta_{kl}\,z^{k}(t_{1})z^{l}(t_{1})\right)\nonumber\\
    &\equiv&-6\,Q_{ij}(t_{1})
\end{eqnarray}
where  $Q_{ij}$ is recognized as the quadrupole moment of the particle per unit mass. 

With this, the Langevin equation of motion can be written compactly in terms of the time derivatives of $Q_{ij}$. 
\begin{eqnarray}\label{Langeqn}
    &&m\delta_{ij}\,\ddot{z}^{j}(t_{1})+\frac{256}{5}\pi^{5}\alpha^{2}z^{j}(t_{1})\Big\{-\delta(0)\,\ddddot{Q}_{ij}(t_{1})\nonumber\\
    &&\hskip 50pt
    +\Big[3\,\delta(t_{1})\ddddot{Q}_{ij}(0)+3\,\dot{\delta}(t_{1})\,\dddot{Q}_{ij}(0)+\ddot{\delta}(t_{1})\,\ddot{Q}_{ij}(0)\Big]\nonumber\\
    &&\hskip 100pt+\frac{1}{2}\big[\theta(t_{1})-\theta(-t_{1})\big]\,\fivedots{Q}_{ij}(t_{1})\Big\}
    =2\,\ddot{\xi}_{ij}(t_{1})\,z^{j}(t_{1})
\end{eqnarray}
The term involving $\delta(0)$ is the renormalization to the interaction term $z^{j}\ddddot{Q}_{ij}$. For a physical system this $\delta(0)$ can be viewed as the regularized one with $\delta_{\Lambda}(0)=\Lambda/\pi$. $\Lambda$ will then be related to the physical scale of the system.

In appearance, there are terms with initial values of $\ddot{Q}_{ij}$, $\dddot{Q}_{ij}$, and $\ddddot{Q}_{ij}$. However, it has been discussed in \cite{HH22} that one  only needs to specify the initial values of $z$ and $\dot{z}$ to determine the dynamics of the system. Note that we have the system with the gravitons and the particle, and everything is causal. After integrating over the graviton degrees of freedom, initial values of the higher time derivatives of the particle coordinate then appear, but they are not necessary in the determination of the particle dynamics. The presence of these initial value terms  is due to the sudden switch-on of the interaction at $t=0$. If one  gradually switch on  the interaction, the weight of these initial value terms will diminish accordingly \cite{FRH11}.

{\section{Evolution operator of the reduced density matrix of a quantum massive particle}
\label{} 

In the last section, we have derived the influence action due to gravitons and have analyzed its effects on the geodesic motion of particles. This is summarized in the resulting Langevin equation and its solutions. The noise effects on the classical geodesics and their congruences were analyzed by the authors in \cite{ChoHu23}. Here, we  extend our considerations to the evolution of \textit{quantum} massive particles in a graviton field necessary for the consideration of quantum information issues such as quantum decoherence and quantum entanglement. We look at their density matrices and their evolution through the quantum master equation. 

The coarse-grained effective action is used to construct the reduced density operator elements of the geodesic deviation $z^{i}(t)$ at time $t$
\begin{align}\label{E:bsiai} 
    \rho(\vec{z},\vec{z}\,';t)=\int d^{3}z_{0}\!\int d^{3}z'_{0}\;\rho(\vec{z}_{0},\vec{z}\,'_{0};0)\int^{\vec{z}}_{\vec{z}_{0}}D\vec{z}_{+}\int^{\vec{z}\,'}_{\vec{z}\,'_{0}}D\vec{z}_{-}\,e^{iS_{CG}[z_{+},z_{-}]}
\end{align}
where $\rho(\vec{z}_{0},\vec{z}\,'_{0};0)$ is the density matrix elements of the initial state of the geodesic deviation $\vec{z}$ at the initial time $t=0$, and $\vec{z}_{0}$, $\vec{z}\,'_{0}$ are allowed initial values of $\vec{z}$.

With these elements of the reduced density operator of $\vec{z}$, we can derive the master equation of $\vec{z}$ using the approach described in~\cite{HM94}. Moreover, we can use the density matrix element \eqref{E:bsiai} to investigate the nonequilibrium evolution of the observables associated with the geodesic deviation because given an operator function $\hat{\mathcal{O}}$ of $\vec{z}$, the corresponding expectation value is formally given by
\begin{equation}
    \langle\hat{\mathcal{O}}(t)\rangle=\operatorname{Tr}\left\{\hat{\rho}(t)\hat{\mathcal{O}}(t)\right\}=\int d^{3}z\int d^{3}z'\;\rho(\vec{z},\vec{z}\,';t)\,\langle\vec{z}\,'\vert\hat{\mathcal{O}}(t)\vert \vec{z}\,\rangle\,.
\end{equation}  
The elements of the density matrix $\rho(\vec{z},\vec{z}\,';t)$ are typically very difficult to evaluate due to the non-linear terms in the influence action $S_{IF}[{z}_{+},{z}_{-}]$, so we need to resort to perturbative techniques in quantum field theory.

First, we start with the reduced density matrix in Eq.~(\ref{E:bsiai}).
\begin{eqnarray}\label{defmaster}
 \rho(\vec{z},\vec{z}\,';t)=\int d^{3}z_{0}\!\int d^{3}z'_{0}\;\rho(\vec{z}_{0},\vec{z}\,'_{0};0){\cal J}(\vec{z},\vec{z}\,';t|\vec{z}_{0},\vec{z}\,'_{0};0)\,,
\end{eqnarray}
where we have defined the evolution operator
\begin{eqnarray}
{\cal J}(\vec{z},\vec{z}\,';t|\vec{z}_{0},\vec{z}\,'_{0};0)=\int^{\vec{z}}_{\vec{z}_{0}}D\vec{z}_{+}\int^{\vec{z}\,'}_{\vec{z}\,'_{0}}D\vec{z}_{-}\,e^{iS_{CG}[z_{+},z_{-}]}
\end{eqnarray}
with the coarse-grained effective action $S_{CG}$ given by Eqs.~(\ref{IFaction2}) to (\ref{SCG}). Note that in the path integral, the initial and final values of $\vec{z}_{+}$ and $\vec{z}_{-}$ are fixed as shown. To evaluate this evolution operator, it is convenient to rewrite $S_{CG}$ in terms of 
\begin{eqnarray}\label{zplusminus}
    \Sigma^{i}(t_{1})=\frac{z_{+}^{i}(t_{1})+z_{-}^{i}(t_{1})}{2},\hskip 50pt \Delta^{i}(t_{1})=z_{+}^{i}(t_{1})-z_{-}^{i}(t_{1}).
\end{eqnarray}
Then, the coarse-grained effective action becomes
\begin{eqnarray}\label{actioncg}
S_{CG}[\vec{\Sigma},\vec{\Delta}]
&=&m\int_{0}^{t}dt_{1}\,\delta_{ij}\left[\dot{\Sigma}^{i}(t_{1})\dot{\Delta}^{j}(t_{1})\right]\nonumber\\
&&+2\int_{0}^{t}dt_{1}\int_{0}^{t}dt_{2}\,\left[\frac{d^{2}}{dt_{1}^{2}}\left(\Delta^{i}(t_{1})\Sigma^{j}(t_{1})\right)\right]D_{ijkl}(t_{1},t_{2})\nonumber\\
&&\hskip 100pt\left[\frac{d^{2}}{dt_{2}^{2}}\left(\Sigma^{k}(t_{2})\Sigma^{l}(t_{2})+\frac{1}{4}\Delta^{k}(t_{2})\Delta^{l}(t_{2})\right)\right]\nonumber\\
&&+2i\int_{0}^{t}dt_{1}\int_{0}^{t}dt_{2}\,\left[\frac{d^{2}}{dt_{1}^{2}}\left(\Delta^{i}(t_{1})\Sigma^{j}(t_{1})\right)\right]N_{ijkl}(t_{1},t_{2})\left[\frac{d^{2}}{dt_{2}^{2}}\left(\Delta^{k}(t_{2})\Sigma^{l}(t_{2})\right)\right]\nonumber\\
\end{eqnarray}
The evolution operator can now be expressed in terms of path integrations over $\vec{\Sigma}$ and $\vec{\Delta}$ as
\begin{eqnarray}
    {\cal J}(\vec{\Sigma},\vec{\Delta};t|\vec{\Sigma}_{0},\vec{\Delta}_{0};0)=\int^{\vec{\Sigma}}_{\vec{\Sigma}_{0}}D\vec{\Sigma}\int^{\vec{\Delta}}_{\vec{\Delta}_{0}}D\vec{\Delta}\;e^{iS_{CG}[\vec{\Sigma},\vec{\Delta}]}
\end{eqnarray}

Since the terms in the influence action are quartic in the field variables, it is not possible to evaluate the path integrations to obtain the exact evolution operator. We shall resort to the perturbation method. To do that, we first consider the free action without the influence action $S_{IF}$ and the corresponding free evolution operator
\begin{eqnarray}\label{freeop}
     {\cal J}_{0}(\vec{\Sigma},\vec{\Delta};t|\vec{\Sigma}_{0},\vec{\Delta}_{0};0)=\int^{\vec{\Sigma}}_{\vec{\Sigma}_{0}}D\vec{\Sigma}\int^{\vec{\Delta}}_{\vec{\Delta}_{0}}D\vec{\Delta}\;e^{i\,m\int_{0}^{t}dt_{1}\,\delta_{ij}\left[\dot{\Sigma}^{i}(t_{1})\dot{\Delta}^{j}(t_{1})\right]}
\end{eqnarray}
To deal with the boundary conditions, we split the functions
\begin{eqnarray}
    \Sigma^{i}=\Sigma_{cl}^{i}+x_{+}^{i},\hskip 50pt 
    \Delta^{i}=\Delta_{cl}^{i}+x_{-}^{i},
\end{eqnarray}
where $\vec{\Sigma}_{cl}(t)$ and $\vec{\Delta}_{cl}(t)$ are classical solutions satisfying the equations of motion
\begin{eqnarray}
    \ddot{\Sigma}_{cl}^{i}=0,\hskip 50pt
    \ddot{\Delta}_{cl}^{i}=0,
\end{eqnarray}
with the boundary conditions
\begin{eqnarray}
   && \Sigma_{cl}^{i}(0)=\Sigma^{i}_{0},\hskip 50pt
    \Delta_{cl}(0)^{i}=\Delta^{i}_{0}\nonumber\\
    && \Sigma_{cl}^{i}(t)=\Sigma^{i},\hskip 65pt
    \Delta_{cl}^{i}(t)=\Delta^{i}\nonumber\\
    &&x_{+}^{i}(0)=x_{-}^{i}(0)=x_{+}^{i}(t)=x_{-}^{i}(t)=0.
\end{eqnarray}
In fact, in this simple case, the classical solutions are simply
\begin{eqnarray}\label{clasols}
    &&\Sigma_{cl}^{i}(t_{1})=\Sigma^{i}\left(\frac{t_{1}}{t}\right)+\Sigma^{i}_{0}\left(\frac{t-t_{1}}{t}\right)\, ,\nonumber\\
    &&\Delta_{cl}^{i}(t_{1})=\Delta^{i}\left(\frac{t_{1}}{t}\right)+\Delta^{i}_{0}\left(\frac{t-t_{1}}{t}\right)
\end{eqnarray}
Using these classical solutions, the free evolution operator in Eq.~(\ref{freeop}) can be readily evalutated.
\begin{eqnarray}\label{J0}
    {\cal J}_{0}(\vec{\Sigma},\vec{\Delta};t|\vec{\Sigma}_{0},\vec{\Delta}_{0};0)&=&e^{i\left(\frac{m}{t}\right)\delta_{ij}\left(\Sigma^{i}-\Sigma^{i}_{0}\right)\left(\Delta^{j}-\Delta^{j}_{0}\right)}\int D\vec{x}_{+}\int D\vec{x}_{-}\,e^{i\,m\int_{0}^{t}dt_{1}\,\delta_{kl}\,\dot{x}^{k}_{+}(t_{1})\dot{x}^{l}_{-}(t_{1})}\;\nonumber\\
    &=&\left(\frac{m}{2\pi t}\right)^{3}e^{i\left(\frac{m}{t}\right)\delta_{ij}\left(\Sigma^{i}-\Sigma^{i}_{0}\right)\left(\Delta^{j}-\Delta^{j}_{0}\right)}
\end{eqnarray}
The path integral over $\vec{x}_{+}$ and $\vec{x}_{-}$ above is just the one for free particles with only kinetic terms. It has been evaluated, for example, in \cite{HM94}.

With the influence action, one can develop a perturbative expansion of the evolution operator by putting in source terms to $\vec{x}_{+}$ and $\vec{x}_{-}$ in the path integrals.
\begin{eqnarray}
&&{\cal J}(\vec{\Sigma},\vec{\Delta};t|\vec{\Sigma}_{0},\vec{\Delta}_{0};0)\nonumber\\
&=&e^{i\left(\frac{m}{t}\right)\delta_{ij}\left(\Sigma^{i}-\Sigma^{i}_{0}\right)\left(\Delta^{j}-\Delta^{j}_{0}\right)}\,e^{iS_{IF}[\Sigma_{cl}+\frac{1}{i}\frac{\delta}{\delta j_{-}},\Delta_{cl}+\frac{1}{i}\frac{\delta}{\delta j_{+}}]}\nonumber\\
&&\ \ \left.\int D\vec{x}_{+}\int D\vec{x}_{-}\,e^{i\int_{0}^{t}dt_{1}\,\delta_{kl}\,[m\dot{x}^{k}_{+}(t_{1})\dot{x}^{l}_{-}(t_{1})+x^{k}_{+}(t_{1})j^{l}_{-}(t_{1})+x^{k}_{-}(t_{1})j^{l}_{+}(t_{1})]}\right|_{j_{+}=j_{-}=0}\nonumber\\
&=& {\cal J}_{0}(\vec{\Sigma},\vec{\Delta};t|\vec{\Sigma}_{0},\vec{\Delta}_{0};0)\nonumber\\
&&\ \ \left.e^{iS_{IF}[\Sigma_{cl}+\frac{1}{i}\frac{\delta}{\delta j_{-}},\Delta_{cl}+\frac{1}{i}\frac{\delta}{\delta j_{+}}]}e^{\frac{i}{m}\int_{0}^{t}dt_{1}\int_{0}^{t}dt_{2}\,\delta_{ij}[j_{+}^{i}(t_{1})G(t_{1},t_{2})j_{-}^{j}(t_{2})]}\right|_{j_{+}=j_{-}=0}
\end{eqnarray}
where we have made use of the Green function
\begin{eqnarray}
    G(t_{1},t_{2})=\theta(t_{1}-t_{2})\left[-\frac{(t-t_{1})\,t_{2}}{t}\right]+\theta(t_{2}-t_{1})\left[-\frac{(t-t_{2})\,t_{1}}{t}\right]
\end{eqnarray}
which satisfies 
\begin{eqnarray}
    \frac{d^{2}}{dt_{1}^{2}}\,G(t_{1},t_{2})=\delta(t_{1}-t_{2})
\end{eqnarray}
with the boundary conditions $G(0,t_{2})=G(t,t_{2})=0$. 

Expanding $e^{iS_{IF}[\Sigma_{cl}+\frac{1}{i}\frac{\delta}{\delta j_{-}},\Delta_{cl}+\frac{1}{i}\frac{\delta}{\delta j_{+}}]}$, one can develop a perturbative series 
\begin{eqnarray} \label{Jxp}
    &&{\cal J}(\vec{\Sigma},\vec{\Delta};t|\vec{\Sigma}_{0},\vec{\Delta}_{0};0)\nonumber\\
    &=&{\cal J}_{0}(\vec{\Sigma},\vec{\Delta};t|\vec{\Sigma}_{0},\vec{\Delta}_{0};0)\left\{1+iS_{IF}\left[\Sigma_{cl}+\frac{1}{i}\frac{\delta}{\delta j_{-}},\Delta_{cl}+\frac{1}{i}\frac{\delta}{\delta j_{+}}\right]+\cdots\right\}\nonumber\\
&&\hskip 40pt\left.e^{\frac{i}{m}\int_{0}^{t}dt_{1}\int_{0}^{t}dt_{2}\,\delta_{ij}[j_{+}^{i}(t_{1})G(t_{1},t_{2})j_{-}^{j}(t_{2})]}\right|_{j_{+}=j_{-}=0}
\nonumber\\
&=&{\cal J}_{0}(\vec{\Sigma},\vec{\Delta};t|\vec{\Sigma}_{0},\vec{\Delta}_{0};0)\left[1+i\delta{\cal A}(\vec{\Sigma},\vec{\Delta};t|\vec{\Sigma}_{0},\vec{\Delta}_{0};0)+\cdots\right]
\end{eqnarray}
Here we have denoted the first order result as $i\delta{\cal A}$ in which we have evaluated the functional derivatives of $\vec{j}_{+}(t_{1})$ and $\vec{j}_{-}(t_{1})$. The result is
\begin{eqnarray}
    i\delta{\cal A}(\vec{\Sigma},\vec{\Delta};t|\vec{\Sigma}_{0},\vec{\Delta}_{0};0)
    &=&2i\int_{0}^{t}dt_{1}\int_{0}^{t}dt_{2}\,D_{ijkl}(t_{1},t_{2})\left(\frac{d^{2}}{dt_{1}^{2}}\right)\left(\frac{d^{2}}{dt_{2}^{2}}\right)\nonumber\\
    &&\ \ \Bigg\{\Delta_{cl}^{i}(t_{1})\Sigma_{cl}^{j}(t_{1})\left[\Sigma_{cl}^{k}(t_{2})\Sigma_{cl}^{l}(t_{2})+\frac{1}{4}\Delta_{cl}^{k}(t_{2})\Delta_{cl}^{l}(t_{2})\right]\nonumber\\
    &&\ \ \ \ 
    -\frac{i}{m}G(t_{1},t_{2})\left[\delta^{ik}\Sigma_{cl}^{j}(t_{1})\Sigma_{cl}^{l}(t_{2})+\delta^{il}\Sigma_{cl}^{j}(t_{1})\Sigma_{cl}^{k}(t_{2})\right]\nonumber\\
    &&\ \ \ \ -\frac{i}{4m}G(t_{1},t_{2})\left[\delta^{jk}\Delta_{cl}^{i}(t_{1})\Delta_{cl}^{l}(t_{2})+\delta^{jl}\Delta_{cl}^{i}(t_{1})\Delta_{cl}^{k}(t_{2})\right]\Bigg\}\nonumber\\
    &&-2\int_{0}^{t}dt_{1}\int_{0}^{t}dt_{2}\,N_{ijkl}(t_{1},t_{2})\left(\frac{d^{2}}{dt_{1}^{2}}\right)\left(\frac{d^{2}}{dt_{2}^{2}}\right)\nonumber\\
    &&\ \ \Bigg\{\Delta_{cl}^{i}(t_{1})\Sigma_{cl}^{j}(t_{1})\Delta_{cl}^{k}(t_{2})\Sigma_{cl}^{l}(t_{2})-\frac{i}{m}G(t_{1},t_{2})\delta^{il}\Sigma_{cl}^{j}(t_{1})\Delta_{cl}^{k}(t_{2})\nonumber\\
    &&\ \ \ \ \ \ \ \
    -\frac{i}{m}G(t_{1},t_{2})\delta^{jk}\Delta_{cl}^{i}(t_{1})\Sigma_{cl}^{l}(t_{2})\Bigg\}
\end{eqnarray}
where we have only retained the connected terms.

In this expression the classical solutions $\Delta_{cl}(t_{1})$ and $\Sigma_{cl}(t_{1})$ are given by Eq.~(\ref{clasols}) in terms of the initial values $\vec{\Delta}_{0}$ and $\vec{\Sigma}_{0}$ and the final values $\vec{\Delta}$ and $\vec{\Sigma}$. In anticipation of using the evolution operator to derive the master equation as indicated in Eq.~(\ref{defmaster}), one will have to integrate over the initial values. To facilitate this operation, it is convenient to turn $\vec{\Delta}_{0}$ and $\vec{\Sigma}_{0}$ into derivatives $\frac{\partial}{\delta\vec{\Sigma}}$ and $\frac{\partial}{\delta\vec{\Delta}}$ acting on ${\cal J}_{0}(\vec{\Sigma},\vec{\Delta};t|\vec{\Sigma}_{0},\vec{\Delta}_{0};0)$. With the expression for the free evolution operator as shown in Eq.~(\ref{J0}), this can be done using the identities
\begin{eqnarray}\label{identities}
    \vec{\Delta}_{0}{\cal J}_{0}(\vec{\Sigma},\vec{\Delta};t|\vec{\Sigma}_{0},\vec{\Delta}_{0};0)&=&\left[\vec{\Delta}+i\left(\frac{t}{m}\right)\frac{\partial}{\partial\vec{\Sigma}}\right]{\cal J}_{0}(\vec{\Sigma},\vec{\Delta};t|\vec{\Sigma}_{0},\vec{\Delta}_{0};0)\\
    \vec{\Sigma}_{0}{\cal J}_{0}(\vec{\Sigma},\vec{\Delta};t|\vec{\Sigma}_{0},\vec{\Delta}_{0};0)&=&\left[\vec{\Sigma}+i\left(\frac{t}{m}\right)\frac{\partial}{\partial\vec{\Delta}}\right]{\cal J}_{0}(\vec{\Sigma},\vec{\Delta};t|\vec{\Sigma}_{0},\vec{\Delta}_{0};0)
\end{eqnarray}
The expressions which are useful in simplifying $\delta{\cal A}{\cal J}_{0}$ in this manner are listed in Appendix \ref{onJ0}. After these substitutions, the expression for $\delta{\cal A}$ is still quite lengthy. It can be further simplified if we contract the expressions with the tensor structure of the noise and the dissipation kernels. Let us write these kernels as
\begin{eqnarray}
    N^{ijkl}(t_{1},t_{2})&=&[2\delta_{ij}\delta_{kl}-3(\delta_{ik}\delta_{jl}+\delta_{il}\delta_{jk})]\,N(t_{1},t_{2})\\
    D^{ijkl}(t_{1},t_{2})&=&[2\delta_{ij}\delta_{kl}-3(\delta_{ik}\delta_{jl}+\delta_{il}\delta_{jk})]\,D(t_{1},t_{2})
\end{eqnarray}
where
\begin{eqnarray}
    N(t_{1},t_{2})&=&-\frac{32\pi^4\alpha^2}{15}\int_{0}^{\infty}dq\,q\,\cos[q(t_{1}-t_{2})]\left(1+\frac{2}{e^{q\beta}-1}\right)\\
    D(t_{1},t_{2})&=&-\frac{64\pi^4\alpha^2}{15}\theta(t_{1}-t_{2})\int_{0}^{\infty}dq\,q\,\sin[q(t_{1}-t_{2})]
\end{eqnarray}
Contracting the tensor structure, the term in $\delta{\cal A}$ involving the noise kernel can be simplified to
\begin{eqnarray}\label{AJnoise}
    &&-2\int_{0}^{t}dt_{1}\int_{0}^{t}dt_{2}\,N_{ijkl}(t_{1},t_{2})\left(\frac{d^{2}}{dt_{1}^{2}}\right)\left(\frac{d^{2}}{dt_{2}^{2}}\right)\nonumber\\
    &&\ \ \Bigg\{\Delta_{cl}^{i}(t_{1})\Sigma_{cl}^{j}(t_{1})\Delta_{cl}^{k}(t_{2})\Sigma_{cl}^{l}(t_{2})-\frac{i}{m}G(t_{1},t_{2})\delta^{il}\Sigma_{cl}^{j}(t_{1})\Delta_{cl}^{k}(t_{2})\nonumber\\
    &&\ \ \ \ \ \ \ \
    -\frac{i}{m}G(t_{1},t_{2})\delta^{jk}\Delta_{cl}^{i}(t_{1})\Sigma_{cl}^{l}(t_{2})\Bigg\}\nonumber\\
    &=&-2\int_{0}^{t}dt_{1}\int_{0}^{t}dt_{2}\,N(t_{1},t_{2})\nonumber\\
    &&\ \ \Bigg\{\left[\frac{120}{m^{2}t^{2}}-\frac{60}{m^{2}t}\left(\frac{d^{2}}{dt_{1}^{2}}\right)\left(\frac{d^{2}}{dt_{2}^{2}}\right)[(t-t_{1})(t-t_{2})G(t_{1},t_{2})]\right]\nonumber\\
    &&\ \ \ \ +i\frac{20}{m}\left(\frac{d^{2}}{dt_{1}^{2}}\right)\left(\frac{d^{2}}{dt_{2}^{2}}\right)G(t_{1},t_{2})\,\Sigma^{i}\Delta_{i}\nonumber\\
    &&\ \ +\left[i\frac{80}{m^{3}t}-i\frac{20}{m^{3}}\left(\frac{d^{2}}{dt_{1}^{2}}\right)\left(\frac{d^{2}}{dt_{2}^{2}}\right)[(t-t_{1})(t-t_{2})G(t_{1},t_{2})]\right]\frac{\partial}{\partial\Delta^{i}}\frac{\partial}{\partial\Sigma_{i}}\nonumber\\
    &&\ \ -\frac{4}{m^{4}}\left(3\frac{\partial}{\partial\Delta^{i}}\frac{\partial}{\partial\Delta_{i}}\frac{\partial}{\partial\Sigma^{j}}\frac{\partial}{\partial\Sigma_{j}}+\frac{\partial}{\partial\Delta^{i}}\frac{\partial}{\partial\Sigma_{i}}\frac{\partial}{\partial\Delta^{j}}\frac{\partial}{\partial\Sigma_{j}}\right)\Bigg\}{\cal J}_{0}(\vec{\Sigma},\vec{\Delta};t|\vec{\Sigma}_{0},\vec{\Delta}_{0};0)
\end{eqnarray}
Similarly, for the term involving the dissipation kernel, we have
\begin{eqnarray}\label{AJdis}
    &&2i\int_{0}^{t}dt_{1}\int_{0}^{t}dt_{2}\,D_{ijkl}(t_{1},t_{2})\left(\frac{d^{2}}{dt_{1}^{2}}\right)\left(\frac{d^{2}}{dt_{2}^{2}}\right)\nonumber\\
    &&\ \ \Bigg\{\Delta_{cl}^{i}(t_{1})\Sigma_{cl}^{j}(t_{1})\left[\Sigma_{cl}^{k}(t_{2})\Sigma_{cl}^{l}(t_{2})+\frac{1}{4}\Delta_{cl}^{k}(t_{2})\Delta_{cl}^{l}(t_{2})\right]\nonumber\\
    &&\ \ \ \ 
    -\frac{i}{m}G(t_{1},t_{2})\left[\delta^{ik}\Sigma_{cl}^{j}(t_{1})\Sigma_{cl}^{l}(t_{2})+\delta^{il}\Sigma_{cl}^{j}(t_{1})\Sigma_{cl}^{k}(t_{2})\right]\nonumber\\
    &&\ \ \ \ -\frac{i}{4m}G(t_{1},t_{2})\left[\delta^{jk}\Delta_{cl}^{i}(t_{1})\Delta_{cl}^{l}(t_{2})+\delta^{jl}\Delta_{cl}^{i}(t_{1})\Delta_{cl}^{k}(t_{2})\right]\Bigg\}
   {\cal J}_{0}(\vec{\Sigma},\vec{\Delta};t|\vec{\Sigma}_{0},\vec{\Delta}_{0};0)\nonumber\\
    &=& 2i\int_{0}^{t}dt_{1}\int_{0}^{t}dt_{2}\,D(t_{1},t_{2})\nonumber\\
    &&\ \ \Bigg\{i\left(\frac{20}{m}\right)\left(\frac{d^{2}}{dt_{1}^{2}}\right)\left(\frac{d^{2}}{dt_{2}^{2}}\right)G(t_{1},t_{2})\left(\Sigma^{i}\Sigma_{i}+\frac{1}{4}\Delta^{i}\Delta_{i}\right)\nonumber\\
    &&\ \ \ \ -\left(\frac{20}{m^{2}}\right)\left(\frac{d^{2}}{dt_{1}^{2}}\right)\left(\frac{d^{2}}{dt_{2}^{2}}\right)\left[(t-t_{1}+t-t_{2})G(t_{1},t_{2})\right]\left(\Sigma^{i}\frac{\partial}{\partial\Delta^{i}}+\frac{1}{4}\Delta^{i}\frac{\partial}{\partial\Sigma^{i}}\right)\nonumber\\
    &&\ \ \ \ +i\left(\frac{20}{m^{3}}\right)\left[\frac{4}{t}-\left(\frac{d^{2}}{dt_{1}^{2}}\right)\left(\frac{d^{2}}{dt_{2}^{2}}\right)((t-t_{1})(t-t_{2})G(t_{1},t_{2}))\right]\left(\frac{\partial}{\partial\Delta^{i}}\frac{\partial}{\partial\Delta_{i}}+\frac{1}{4}\frac{\partial}{\partial\Sigma^{i}}\frac{\partial}{\partial\Sigma_{i}}\right)\nonumber\\
    &&\ \ \ \ -\left(\frac{16}{m^{4}}\right)\frac{\partial}{\partial\Sigma^{i}}\frac{\partial}{\partial\Delta_{i}}\left(\frac{\partial}{\partial\Delta^{j}}\frac{\partial}{\partial\Delta_{j}}+\frac{1}{4}\frac{\partial}{\partial\Sigma^{j}}\frac{\partial}{\partial\Sigma_{j}}\right)\Bigg\}{\cal J}_{0}(\vec{\Sigma},\vec{\Delta};t|\vec{\Sigma}_{0},\vec{\Delta}_{0};0)
\end{eqnarray}

Combining the results in Eqs.~(\ref{AJnoise}) and (\ref{AJdis}), we obtain the lowest order contribution to the evolution operator
\begin{eqnarray}\label{opexp}
    {\cal J}(\vec{\Sigma},\vec{\Delta};t|\vec{\Sigma}_{0},\vec{\Delta}_{0};0)=\left[1+i\delta{\cal A}(\vec{\Sigma},\vec{\Delta};t|\vec{\Sigma}_{0},\vec{\Delta}_{0};0)+\cdots\right]{\cal J}_{0}(\vec{\Sigma},\vec{\Delta};t|\vec{\Sigma}_{0},\vec{\Delta}_{0};0)
\end{eqnarray}
with
\begin{eqnarray}\label{deltaA}
    &&i\delta{\cal A}(\vec{\Sigma},\vec{\Delta};t|\vec{\Sigma}_{0},\vec{\Delta}_{0};0){\cal J}_{0}(\vec{\Sigma},\vec{\Delta};t|\vec{\Sigma}_{0},\vec{\Delta}_{0};0)\nonumber\\
    &=&\Bigg\{n_{1}(t)+in_{2}(t)\Sigma^{i}\Delta_{i}+in_{3}(t)\frac{\partial}{\partial\Delta^{i}}\frac{\partial}{\partial\Sigma_{i}}\nonumber\\
    &&\ \ 
    +n_{4}(t)\left(3\frac{\partial}{\partial\Delta^{i}}\frac{\partial}{\partial\Delta_{i}}\frac{\partial}{\partial\Sigma^{j}}\frac{\partial}{\partial\Sigma_{j}}+\frac{\partial}{\partial\Delta^{i}}\frac{\partial}{\partial\Sigma_{i}}\frac{\partial}{\partial\Delta^{j}}\frac{\partial}{\partial\Sigma_{j}}\right)\nonumber\\
    &&\ \ +d_{1}(t)\left(\Sigma^{i}\Sigma_{i}+\frac{1}{4}\Delta^{i}\Delta_{i}\right)+id_{2}(t)\left(\Sigma^{i}\frac{\partial}{\partial\Delta^{i}}+\frac{1}{4}\Delta^{i}\frac{\partial}{\partial\Sigma^{i}}\right)\nonumber\\
    &&\ \ 
    +d_{3}(t)\left(\frac{\partial}{\partial\Delta^{i}}\frac{\partial}{\partial\Delta_{i}}+\frac{1}{4}\frac{\partial}{\partial\Sigma^{i}}\frac{\partial}{\partial\Sigma_{i}}\right)\nonumber\\
    &&\ \ +id_{4}(t)\frac{\partial}{\partial\Sigma^{i}}\frac{\partial}{\partial\Delta_{i}}\left(\frac{\partial}{\partial\Delta^{j}}\frac{\partial}{\partial\Delta_{j}}+\frac{1}{4}\frac{\partial}{\partial\Sigma^{j}}\frac{\partial}{\partial\Sigma_{j}}\right)\Bigg\}{\cal J}_{0}(\vec{\Sigma},\vec{\Delta};t|\vec{\Sigma}_{0},\vec{\Delta}_{0};0)
\end{eqnarray}
where
\begin{eqnarray}
    n_{1}(t)&=&-\left(\frac{120}{m^{2}}\right)\int_{0}^{t}dt_{1}\int_{0}^{t}dt_{2}\,N(t_{1},t_{2})\left[\frac{2}{t^{2}}-\frac{1}{t}\left(\frac{d^{2}}{dt_{1}^{2}}\right)\left(\frac{d^{2}}{dt_{2}^{2}}\right)[(t-t_{1})(t-t_{2})G(t_{1},t_{2})]\right]\nonumber\\ \label{n1} \\
    n_{2}(t)&=&-\left(\frac{40}{m}\right)\int_{0}^{t}dt_{1}\int_{0}^{t}dt_{2}\,N(t_{1},t_{2})\left(\frac{d^{2}}{dt_{1}^{2}}\right)\left(\frac{d^{2}}{dt_{2}^{2}}\right)G(t_{1},t_{2})\\
    n_{3}(t)&=&-\left(\frac{40}{m^{3}}\right)\int_{0}^{t}dt_{1}\int_{0}^{t}dt_{2}\,N(t_{1},t_{2})
    \left[\frac{4}{t}-\left(\frac{d^{2}}{dt_{1}^{2}}\right)\left(\frac{d^{2}}{dt_{2}^{2}}\right)[(t-t_{1})(t-t_{2})G(t_{1},t_{2})]\right]\nonumber\\ \\
    n_{4}(t)&=&\frac{8}{m^{4}}\int_{0}^{t}dt_{1}\int_{0}^{t}dt_{2}\,N(t_{1},t_{2})\\
    d_{1}(t)&=&-\left(\frac{40}{m}\right)\int_{0}^{t}dt_{1}\int_{0}^{t}dt_{2}\,D(t_{1},t_{2})\left(\frac{d^{2}}{dt_{1}^{2}}\right)\left(\frac{d^{2}}{dt_{2}^{2}}\right)G(t_{1},t_{2})\label{d1}\\
    d_{2}(t)&=&-\left(\frac{40}{m^{2}}\right)\int_{0}^{t}dt_{1}\int_{0}^{t}dt_{2}\,D(t_{1},t_{2})\left(\frac{d^{2}}{dt_{1}^{2}}\right)\left(\frac{d^{2}}{dt_{2}^{2}}\right)[(t-t_{1}+t-t_{2})G(t_{1},t_{2})]\\
    d_{3}(t)&=&-\left(\frac{40}{m^{3}}\right)\int_{0}^{t}dt_{1}\int_{0}^{t}dt_{2}\,D(t_{1},t_{2})
    \left[\frac{4}{t}-\left(\frac{d^{2}}{dt_{1}^{2}}\right)\left(\frac{d^{2}}{dt_{2}^{2}}\right)[(t-t_{1})(t-t_{2})G(t_{1},t_{2})]\right]\nonumber\\ \\
    d_{4}(t)&=&-\frac{32}{m^{4}}\int_{0}^{t}dt_{1}\int_{0}^{t}dt_{2}\,D(t_{1},t_{2}) \label{d4}
\end{eqnarray}
With this evolution operator it is now possible to derive the master equation for the density matrix of the particle from Eq.~(\ref{defmaster}). This will be worked out in the next section. Note that the coefficient functions from Eqs.~(\ref{n1}) to (\ref{d4}) are in general divergent and need  to be regularized. The kernels $N(t_{1},t_{2})$ and $D(t_{1},t_{2})$ involve integrations over $q$. As we have discussed in the last section, an ultraviolet cutoff scale $\Lambda$ has to be implemented to render these $q$-integrals finite. $\Lambda$ represents a cutoff scale for the gravitons in the environmental degrees of freedom. On the other hand, the $t$-derivatives on the Green function $G(t_{1},t_{2})$ will also produce divergent quantities like the delta function and its derivatives. These divergent quantities can also be regularized in the same way as discussed in Appendix A by introducing another cutoff scale $\lambda$, for example,
\begin{eqnarray}\label{deltalambda}
    \delta_{\lambda}(t_{1}-t_{2})=\frac{\sin[\lambda (t_{1}-t_{2})]}{\pi(t_{1}-t_{2})}
\end{eqnarray}
This $\lambda$ represents a cutoff scale for the quantum particles, and we shall assume that $\Lambda\gg\lambda$. In our subsequent discussions of the master equation in the next section, we shall also need to consider the low temperature ($\Lambda\gg\lambda\gg 1/\beta$) and the high temperature ($1/\beta\gg\Lambda\gg\lambda$) expansions. The coefficient functions of the evolution operator as well as those in the master equation under these expansions will be given in more detail in Appendix C.


{\section{non-Markovian master equation for quantum masses in a graviton field}

To construct the master equation of the reduced density matrix of the quantum particle, we need to calculate the time derivative of the evolution operator. From Eq.~(\ref{opexp}),
\begin{eqnarray}
  \dot{\cal J}=\left(1+i\delta{\cal A}\right)\dot{\cal J}_{0}+i\dot{\delta{\cal A}}\,{\cal J}_{0}+\cdots
\end{eqnarray}
From the result of ${\cal J}_{0}$ in Eq.~(\ref{J0}), we have
\begin{eqnarray}
    \dot{\cal J}_{0}=\frac{i}{m}\delta^{ij}\frac{\partial}{\partial\Sigma^{i}}\frac{\partial}{\partial\Delta^{j}}{\cal J}_{0}
\end{eqnarray}
and 
\begin{eqnarray}\label{tderJ}
    \dot{\cal J}&=&\left(1+i\delta{\cal A}\right)\left(\frac{i}{m}\delta^{ij}\frac{\partial}{\partial\Sigma^{i}}\frac{\partial}{\partial\Delta^{j}}{\cal J}_{0}\right)+i\dot{\delta{\cal A}}\,{\cal J}_{0}+\cdots\nonumber\\
    &=&\frac{i}{m}\delta^{ij}\frac{\partial}{\partial\Sigma^{i}}\frac{\partial}{\partial\Delta^{j}}{\cal J}-\left[\frac{i}{m}\delta^{ij}\frac{\partial}{\partial\Sigma^{i}}\frac{\partial}{\partial\Delta^{j}},i\delta{\cal A}\right]{\cal J}+i\dot{\delta{\cal A}}\,{\cal J}+\cdots
\end{eqnarray}
where in the second equality we have replaced ${\cal J}_{0}$ by ${\cal J}$ which is acceptable up to the first order that we are working with. The commutator can be readily evaluated from the form of the operator $i\delta{\cal A}$ in Eq.~(\ref{deltaA}):
\begin{eqnarray}
    &&\left[\frac{i}{m}\delta^{ij}\frac{\partial}{\partial\Sigma^{i}}\frac{\partial}{\partial\Delta^{j}},i\delta{\cal A}\right]\nonumber\\
    &=&-\left(\frac{1}{m}\right)n_{2}(t)\left(3+\Sigma^{i}\frac{\partial}{\partial\Sigma^{i}}+\Delta^{i}\frac{\partial}{\partial\Delta^{i}}\right)+\left(\frac{i}{m}\right)d_{1}(t)\left(2\Sigma^{i}\frac{\partial}{\partial\Delta^{i}}+\frac{1}{2}\Delta^{i}\frac{\partial}{\partial\Sigma^{i}}\right)\nonumber\\
    &&\ \ -\left(\frac{1}{m}\right)d_{2}(t)\left(\frac{\partial}{\partial\Delta^{i}}\frac{\partial}{\partial\Delta_{i}}+\frac{1}{4}\frac{\partial}{\partial\Sigma^{i}}\frac{\partial}{\partial\Sigma_{i}}\right)
\end{eqnarray}
Putting this into Eq.~(\ref{tderJ}) we can readily obtain the time derivative of the evolution operator. Since the reduced density matrix of the quantum particle is given by Eq.~(\ref{defmaster}), the time derivative of the density matrix is therefore directly related to the time derivative of the evolution operator. In this way, the master equation for the reduced density matrix of the quantum particle is just
\begin{eqnarray}\label{mastereqn1}
    \frac{\partial}{\partial t}\,\rho(\vec{\Sigma},\vec{\Delta};t)
    &=&\int d^{3}\Sigma_{0}\int d^{3}\Delta_{0}\ \dot{{\cal J}}(\vec{\Sigma},\vec{\Delta};t|\vec{\Sigma}_{0},\vec{\Delta}_{0};0)\,\rho(\vec{\Sigma},\vec{\Delta};t)
    \nonumber\\
    &=&\Bigg[\frac{i}{m_{\rm ren}(t)}\frac{\partial}{\partial\Sigma^{i}}\frac{\partial}{\partial\Delta_{i}}+N_{1}(t)+iN_{2}(t)\Sigma^{i}\Delta_{i}\nonumber\\
    &&\ \ +D_{1}(t)\left(\Sigma^{i}\Sigma_{i}+\frac{1}{4}\Delta^{i}\Delta_{i}\right)+iD_{2}(t)\left(\Sigma^{i}\frac{\partial}{\partial\Delta^{i}}+\frac{1}{4}\Delta^{i}\frac{\partial}{\partial\Sigma^{i}}\right)\nonumber\\
    &&\ \ 
    +N_{3}(t)\left(\Sigma^{i}\frac{\partial}{\partial\Sigma^{i}}+\Delta^{i}\frac{\partial}{\partial\Delta^{i}}\right)+D_{3}(t)\left(\frac{\partial}{\partial\Delta^{i}}\frac{\partial}{\partial\Delta_{i}}+\frac{1}{4}\frac{\partial}{\partial\Sigma^{i}}\frac{\partial}{\partial\Sigma_{i}}\right)\nonumber\\
    &&\ \ 
    +N_{4}(t)\left(3\frac{\partial}{\partial\Delta^{i}}\frac{\partial}{\partial\Delta_{i}}\frac{\partial}{\partial\Sigma^{j}}\frac{\partial}{\partial\Sigma_{j}}+\frac{\partial}{\partial\Delta^{i}}\frac{\partial}{\partial\Sigma_{i}}\frac{\partial}{\partial\Delta^{j}}\frac{\partial}{\partial\Sigma_{j}}\right)\nonumber\\
    &&\ \ +iD_{4}(t)\frac{\partial}{\partial\Sigma^{i}}\frac{\partial}{\partial\Delta_{i}}\left(\frac{\partial}{\partial\Delta^{j}}\frac{\partial}{\partial\Delta_{j}}+\frac{1}{4}\frac{\partial}{\partial\Sigma^{j}}\frac{\partial}{\partial\Sigma_{j}}\right)\Bigg]\,\rho(\vec{\Sigma},\vec{\Delta};t)
\end{eqnarray}
where
\begin{eqnarray}\label{mastercof}
    &&\frac{1}{m_{\rm ren}}=\frac{1}{m}+\dot{n}_{3}(t)\hskip 10pt ; \hskip 10pt N_{1}(t)=\dot{n}_{1}(t)+\frac{3}{m}n_{2}(t)\hskip 10pt ;\hskip 10pt N_{2}(t)=\dot{n}_{2}(t)\nonumber\\
    &&D_{1}(t)=\dot{d}_{1}(t)\hskip 10pt ;\hskip 10pt D_{2}(t)=\dot{d}_{2}(t)-\frac{2}{m}d_{1}(t)\hskip 10pt ;\hskip 10pt N_{3}(t)=\frac{1}{m}n_{2}(t)\nonumber\\
    &&D_{3}(t)=\dot{d}_{3}(t)+\frac{1}{m}d_{2}(t)\hskip 10pt ; \hskip 10pt N_{4}(t)=\dot{n}_{4}(t)\hskip 10pt ;\hskip 10pt D_{4}(t)=\dot{d}_{4}(t)
\end{eqnarray}
These coefficient functions are given by the coefficients of the evolution operator and their time derivatives. Therefore, we can develop their low temperature and high temperature expansions which are again listed in Appendix C.

One can express this master equation in an alternative form which is perhaps more transparent for understanding the meanings of various terms in the equation. Since 
\begin{eqnarray}
    \vec{\Sigma}=\frac{1}{2}(\vec{z}+\vec{z}\,')\hskip 10pt ;\hskip 10pt \vec{\Delta}=\vec{z}-\vec{z}\,'
\end{eqnarray}
this will give
\begin{eqnarray}
    \frac{\partial}{\partial\vec{\Sigma}}=\frac{\partial}{\partial \vec{z}}+\frac{\partial}{\partial \vec{z}\,'}\hskip 10pt ;\hskip 10pt \frac{\partial}{\partial\vec{\Delta}}=\frac{1}{2}\left(\frac{\partial}{\partial \vec{z}}-\frac{\partial}{\partial \vec{z}\,'}\right)
\end{eqnarray}
In terms of $\vec{z}$ and $\vec{z}\,'$, the master equation above can be rewritten as
\begin{eqnarray}\label{mastereqn2}
    &&\frac{\partial}{\partial t}\rho(\vec{z},\vec{z}\,',t)\nonumber\\
    &=&\Bigg\{\frac{i}{m_{\rm ren}(t)}\left(\frac{1}{2}\right)\left(\frac{\partial}{\partial\vec{z}}\cdot\frac{\partial}{\partial\vec{z}}-\frac{\partial}{\partial\vec{z}\,'}\cdot\frac{\partial}{\partial\vec{z}\,'}\right)+N_{1}(t)+iN_{2}(t)\left(\frac{1}{2}\right)(\vec{z}\,^{2}-\vec{z}\,'^{2})\nonumber\\
    &&\ \ +D_{1}(t)\left(\frac{1}{2}\right)(\vec{z}\,^{2}+\vec{z}\,'^{2})+iD_{2}(t)\left(\frac{1}{2}\right)\left(\vec{z}\cdot\frac{\partial}{\partial\vec{z}}-\vec{z}\,'\cdot\frac{\partial}{\partial\vec{z}\,'}\right)\nonumber\\
    &&\ \ +N_{3}(t)\left(\frac{1}{2}\right)\left(\vec{z}\cdot\frac{\partial}{\partial\vec{z}}+\vec{z}\,'\cdot\frac{\partial}{\partial\vec{z}\,'}\right)+D_{3}(t)\left(\frac{1}{2}\right)\left(\frac{\partial}{\partial\vec{z}}\cdot\frac{\partial}{\partial\vec{z}}+\frac{\partial}{\partial\vec{z}\,'}\cdot\frac{\partial}{\partial\vec{z}\,'}\right)\nonumber\\
    &&\ \ +N_{4}(t)\left(\frac{1}{4}\right)\Bigg[3\left(\frac{\partial}{\partial\vec{z}}\cdot\frac{\partial}{\partial\vec{z}}+2\,\frac{\partial}{\partial\vec{z}}\cdot\frac{\partial}{\partial\vec{z}\,'}+\frac{\partial}{\partial\vec{z}\,'}\cdot\frac{\partial}{\partial\vec{z}\,'}\right)\nonumber\\
    &&\hskip 80pt\left(\frac{\partial}{\partial\vec{z}}\cdot\frac{\partial}{\partial\vec{z}}-2\,\frac{\partial}{\partial\vec{z}}\cdot\frac{\partial}{\partial\vec{z}\,'}+\frac{\partial}{\partial\vec{z}\,'}\cdot\frac{\partial}{\partial\vec{z}\,'}\right)+\left(\frac{\partial}{\partial\vec{z}}\cdot\frac{\partial}{\partial\vec{z}}-\frac{\partial}{\partial\vec{z}\,'}\cdot\frac{\partial}{\partial\vec{z}\,'}\right)^{2}\Bigg]\nonumber\\
    &&\ \ +iD_{4}(t)\left(\frac{1}{4}\right)\left(\frac{\partial}{\partial\vec{z}}\cdot\frac{\partial}{\partial\vec{z}}-\frac{\partial}{\partial\vec{z}\,'}\cdot\frac{\partial}{\partial\vec{z}\,'}\right)^{2}\Bigg\}\,\rho(\vec{z},\vec{z}\,',t)
\end{eqnarray}

We can see that all terms other than the last two involving $N_{4}(t)$ and $D_{4}(t)$ can be divided into terms depending separately on $\vec{z}$ and $\vec{z}\,'$. Explicitly, we can write
\begin{eqnarray} \label{master1}
    \frac{\partial}{\partial t}\,\rho(\vec{z},\vec{z}\,',t)
    &=&\Bigg\{\bigg[\left(\frac{i}{m_{\rm ren}(t)}+D_{3}(t)\right)\left(\frac{1}{2}\right)\left(\frac{\partial}{\partial\vec{z}}\cdot\frac{\partial}{\partial\vec{z}}\right)+\left(iN_{2}(t)+D_{1}(t)\right)\left(\frac{1}{2}\right)\vec{z}\,^{2}\nonumber\\
    &&\ \ \ \ +(iD_{2}(t)+N_{3}(t))\left(\frac{1}{2}\right)\left(\vec{z}\cdot\frac{\partial}{\partial\vec{z}}\right)+\frac{1}{2}N_{1}(t)\bigg]\nonumber\\
    &&-\bigg[\left(\frac{i}{m_{\rm ren}(t)}-D_{3}(t)\right)\left(\frac{1}{2}\right)\left(\frac{\partial}{\partial\vec{z}\,'}\cdot\frac{\partial}{\partial\vec{z}\,'}\right)+\left(iN_{2}(t)-D_{1}(t)\right)\left(\frac{1}{2}\right)\vec{z}\,'^{2}\nonumber\\
    &&\ \ \ \ +(iD_{2}(t)-N_{3}(t))\left(\frac{1}{2}\right)\left(\vec{z}\,'\cdot\frac{\partial}{\partial\vec{z}\,'}\right)-\frac{1}{2}N_{1}(t)\bigg]\nonumber\\
    &&+N_{4}(t)\cdots+iD_{4}(t)\cdots\Bigg\}\,\rho(\vec{z},\vec{z}\,',t)
\end{eqnarray}
The terms in the first and the second square brackets are at most quadratic in $\vec{z}$ (or $\vec{z}\,'$) and $\partial/\partial\vec{z}$ (or $\partial/\partial\vec{z}\,'$). They can thus be regarded as renormalization contributions to the ``free action" of $\vec{z}$ and $\vec{z}\,'$. The non-trivial dynamics should therefore reside mainly on the interaction terms coming from the noise kernel (the $N_{4}(t)$ term) and the dissipation kernel (the $D_{4}(t)$ term).

The master equation of a quantum system interacting with the graviton environment has been considered previously in \cite{AHGraDec} and \cite{Blencowe}. In \cite{AHGraDec}, the authors start with a massive scalar field interacting with the gravitons. They have also adopted the perturbation method to derive the master to the first order in the Newton constant $G$. Subsequently, a restriction to the one-particle subspace of the scalar field is performed. The master equation they obtain in Eq.~(52) of \cite{AHGraDec} is basically equivalent to what we have here. However, in the next equation (Eq.~(53)) in their paper, the authors have taken a high temperature limit to simplify the equation without specifying it. Using this high temperature or Markovian form, they have analyzed the decoherence phenomenon in the case with one spatial dimension. 

In \cite{Blencowe}, the author again considers the interaction between a massive scalar field and gravitons. Nevertheless, the tensor structure of the graviton has been neglected and it has been treated just like a minimally coupled scalar field. Hence, the master equation obtained in \cite{Blencowe} is not complete. In its subsequent analysis of the decoherence phenomenon, the high temperature limit, which is Markovian as well, has been adopted. 

Our work here deals with the master equation of a quantum particle interacting with the graviton in the general temperature environment. The terms involving $N_{4}(t)$ and $D_{4}(t)$, with four $\vec{z}$ or $\vec{z}\,'$ derivatives indicating the product of four momentum terms or that of two energy terms, are similar to those terms in the master equations derived in \cite{AHGraDec} and \cite{Blencowe}. Yet we have kept the temperature dependence of the coefficient function $N_{4}(t)$ general. In the following subsection, we shall elaborate on the low and high temperature limits in a more detailed fashion. The high temperature results are consistent with those in \cite{AHGraDec} and \cite{Blencowe}. On the other hand, the low temperature results show novel non-Markovian behaviors of the master equation.


\subsection{Fokker-Planck  equation for the reduced Wigner function}
From the form of the master equation for a quantum particle under the influence of gravitons as given in Eq.~(\ref{mastereqn1}), it is straightforward to derive the FP equation for the reduced Wigner distribution function. The Wigner function is defined by
\begin{eqnarray}
    W(\vec{\Sigma},\vec{p},t)=\int d^{3}\!\Delta\ e^{-i\vec{p}\cdot\vec{\Delta}}\, \rho(\vec{\Sigma},\vec{\Delta},t)
\end{eqnarray}
as a Fourier transform of the density matrix, ditto for the reduced Wigner function in relation to the reduced density matrix. From Eq.~(\ref{mastereqn1}) for the reduced density matrix, we obtain the corresponding equation for the reduced Wigner distribution function in the form
\begin{eqnarray}\label{FPeqn}
    \frac{\partial}{\partial t}\,W(\vec{\Sigma},\vec{p},t)
    &=&\Bigg\{-\frac{1}{m_{\rm ren}(t)}\,\vec{p}\cdot\frac{\partial}{\partial\vec{\Sigma}}+N_{1}(t)-N_{2}(t)\,\vec{\Sigma}\cdot\frac{\partial}{\partial\vec{p}}\nonumber\\
    &&\ \ \ \ +D_{1}(t)\left(\vec{\Sigma}\cdot\vec{\Sigma}-\frac{1}{4}\frac{\partial}{\partial\vec{p}}\cdot\frac{\partial}{\partial\vec{p}}\right)-D_{2}(t)\left(\vec{\Sigma}\cdot\vec{p}+\frac{1}{4}\frac{\partial}{\partial\vec{\Sigma}}\cdot\frac{\partial}{\partial\vec{p}}\right)\nonumber\\
    &&\ \ \ \ 
    +N_{3}(t)\left(\vec{\Sigma}\cdot\frac{\partial}{\partial\vec{\Sigma}}-\frac{\partial}{\partial\vec{p}}\cdot\vec{p}\right)+D_{3}(t)\left(-\vec{p}\cdot\vec{p}+\frac{1}{4}\frac{\partial}{\partial\vec{\Sigma}}\frac{\partial}{\partial\vec{\Sigma}}\right)\nonumber\\
    &&\ \ \ \ 
    -N_{4}(t)\left[\,3\,(\vec{p}\cdot\vec{p})\cdot\left(\frac{\partial}{\partial\vec{\Sigma}}\frac{\partial}{\partial\vec{\Sigma}}\right)+\left(\vec{p}\cdot\frac{\partial}{\partial\vec{\Sigma}}\right)^{2}\right]\nonumber\\
    &&\ \ \ \ +D_{4}(t)\left[\left(\vec{p}\cdot\frac{\partial}{\partial\vec{\Sigma}}\right)\left(\vec{p}\cdot\vec{p}-\frac{\partial}{\partial\vec{\Sigma}}\cdot\frac{\partial}{\partial\vec{\Sigma}}\right)\right]\Bigg\}\,W(\vec{\Sigma},\vec{p};t)
\end{eqnarray}
This FP equation is very useful in the analysis of quantum evolution in the phase space. As exemplified in \cite{PHZ93}, both the reduction of wavefunctions and the phenomenon of gravitational decoherence can be studied through this equation in a transparent graphical manner.

\subsection{Markovian limit at high temperature}
As we have discussed in Section IIA, the dissipation kernel of the gravitons can be expressed as the time derivative of a delta function as given in Eq.~(\ref{disfinal}). Moreover, in the leading order of the high temperature limit with $1/\beta\gg\Lambda\gg\lambda$, the noise kernel can also be expressed as being proportional to a delta function. Therefore, we obtain the Markovian limit in which the dissipation kernel is local and the noise kernel corresponds to one with white noise. 

Let us now examine the master equation in this limit. As we have discussed above, we should concentrate on the terms with four derivatives, that is, the terms with coefficient functions $N_{4}(t)$ and $D_{4}(t)$. The high temperature limit of these coefficient functions are given in Appendix C. For $1/\beta\gg\Lambda\gg\lambda$,
\begin{eqnarray}
    N_{4}(t)\sim-\frac{512\pi^{5}\alpha^{2}}{15m^{4}\beta}\ \ ;\ \ D_{4}(t)\sim\frac{2048\pi^{4}\alpha^{2}\Lambda}{15m^{4}}
\end{eqnarray}
Note that $D_{4}(t)$ is independent of temperature. In this limit, the master equation becomes
\begin{eqnarray}
    &&\frac{\partial}{\partial t}\rho(\vec{z},\vec{z}\,',t)\nonumber\\
    &=&\cdots-\left(\frac{512\pi^{5}\alpha^{2}}{15m^{4}\beta}\right)\left(\frac{1}{4}\right)\Bigg[3\left(\frac{\partial}{\partial\vec{z}}\cdot\frac{\partial}{\partial\vec{z}}+2\,\frac{\partial}{\partial\vec{z}}\cdot\frac{\partial}{\partial\vec{z}\,'}+\frac{\partial}{\partial\vec{z}\,'}\cdot\frac{\partial}{\partial\vec{z}\,'}\right)\nonumber\\
    &&\hskip 80pt\left(\frac{\partial}{\partial\vec{z}}\cdot\frac{\partial}{\partial\vec{z}}-2\,\frac{\partial}{\partial\vec{z}}\cdot\frac{\partial}{\partial\vec{z}\,'}+\frac{\partial}{\partial\vec{z}\,'}\cdot\frac{\partial}{\partial\vec{z}\,'}\right)+\left(\frac{\partial}{\partial\vec{z}}\cdot\frac{\partial}{\partial\vec{z}}-\frac{\partial}{\partial\vec{z}\,'}\cdot\frac{\partial}{\partial\vec{z}\,'}\right)^{2}\Bigg]\nonumber\\
    &&\ \ +i\left(\frac{2048\pi^{5}\alpha^{2}\Lambda}{15m^{4}}\right)\left(\frac{1}{4}\right)\left(\frac{\partial}{\partial\vec{z}}\cdot\frac{\partial}{\partial\vec{z}}-\frac{\partial}{\partial\vec{z}\,'}\cdot\frac{\partial}{\partial\vec{z}\,'}\right)^{2}\Bigg\}\,\rho(\vec{z},\vec{z}\,',t)
\end{eqnarray}
As emphasized above, the limit we are taking satisfies the relation $1/\beta\gg\Lambda$. Hence, we can see that $N_{4}(t)\gg D_{4}(t)$. The coefficient coming from the noise kernel dominates over the one from the dissipation kernel. 

From the form of the master equation above, it is apparent that the $N_{4}(t)$ terms leads to decoherence effect on the reduced density matrix $\rho(\vec{z},\vec{z}\,',t)$. The decoherence time scale due to gravitons can therefore be estimated to be
\begin{eqnarray}\label{dectime}
    t_{dec}^{-1}\sim\left(\frac{512\pi^{5}\alpha^{2}}{15m^{4}\beta}\right)\left[3\,\vec{\Sigma}_{p}^{2}\vec{\Delta}_{p}^{2}+(\vec{\Sigma}_{p}\cdot\vec{\Delta}_{p})^{2}\right]
\end{eqnarray}
in the momentum representation with $\vec{\Sigma_{p}}=(\vec{p}+\vec{p}\,')$ and $\vec{\Delta_{p}}=(\vec{p}-\vec{p}\,')/2$. We see that $t_{dec}^{-1}$ is proportional to $1/\beta$ or the temperature $T$, and also to momentum to the fourth power (or energy to the second power). As we have mentioned above, this result is consistent with the Markovian limit of the decoherence time shown in \cite{AHGraDec} and \cite{Blencowe}.

\subsection{Non-Markovian low temperature limit}
It is also interesting to look at the master equation in Eq.~(\ref{mastereqn2}) in the low temperature limit. Again, we concentrate on the $N_{4}(t)$ and $D_{4}(t)$ interaction terms. From Appendix C, we have the low temperature expansions for these coefficient functions. Then, the master equation can be expressed as
\begin{eqnarray}
    &&\frac{\partial}{\partial t}\rho(\vec{z},\vec{z}\,',t)\nonumber\\
    &=&\cdots-\left(\frac{512\pi^{4}\alpha^{2}}{15m^{4}}\right)\left[\frac{1}{t}+\frac{\pi^{2}t}{3\beta^{2}}\right]\left(\frac{1}{4}\right)\Bigg[3\left(\frac{\partial}{\partial\vec{z}}\cdot\frac{\partial}{\partial\vec{z}}+2\,\frac{\partial}{\partial\vec{z}}\cdot\frac{\partial}{\partial\vec{z}\,'}+\frac{\partial}{\partial\vec{z}\,'}\cdot\frac{\partial}{\partial\vec{z}\,'}\right)\nonumber\\
    &&\hskip 80pt\left(\frac{\partial}{\partial\vec{z}}\cdot\frac{\partial}{\partial\vec{z}}-2\,\frac{\partial}{\partial\vec{z}}\cdot\frac{\partial}{\partial\vec{z}\,'}+\frac{\partial}{\partial\vec{z}\,'}\cdot\frac{\partial}{\partial\vec{z}\,'}\right)+\left(\frac{\partial}{\partial\vec{z}}\cdot\frac{\partial}{\partial\vec{z}}-\frac{\partial}{\partial\vec{z}\,'}\cdot\frac{\partial}{\partial\vec{z}\,'}\right)^{2}\Bigg]\nonumber\\
    &&\ \ +i\left(\frac{2048\pi^{5}\alpha^{2}\Lambda}{15m^{4}}\right)\left(\frac{1}{4}\right)\left(\frac{\partial}{\partial\vec{z}}\cdot\frac{\partial}{\partial\vec{z}}-\frac{\partial}{\partial\vec{z}\,'}\cdot\frac{\partial}{\partial\vec{z}\,'}\right)^{2}\Bigg\}\,\rho(\vec{z},\vec{z}\,',t)
\end{eqnarray}
In this low temperature expansion, we have $\Lambda\gg\lambda\gg 1/t\gg 1/\beta$. First, we consider the case with zero temperature. The $D_{4}(t)$ term, which signifies the dissipation effect, is independent of temperature and again proportional to $\Lambda$. Hence, the dissipation effect is quite substantial and long lasting, as expected. On the other hand, the $N_{4}(t)$ term is proportional to $1/t$, which signifies that the time dependence of decoherence effect is no longer exponential. Rewrite the master equation emphasizing the noise $N_{4}(t)$ term in the momentum representation. 
\begin{eqnarray}\label{lowmaster}
    \frac{\partial}{\partial t}\,\rho(\vec{p},\vec{p}\,',t)=\cdots-\left(\frac{a}{t}\right)\,\rho(\vec{p},\vec{p}\,',t)
\end{eqnarray}
where
\begin{eqnarray}
    a=\left(\frac{512\pi^{4}\alpha^{2}}{15m^{4}}\right)\left[3\,\vec{\Sigma}_{p}^{2}\vec{\Delta}_{p}^{2}+(\vec{\Sigma}_{p}\cdot\vec{\Delta}_{p})^{2}\right]
\end{eqnarray}
Solving this equation, we have $\rho(\vec{p},\vec{p}\,',t)\sim t^{-a}$. However, $a$ is proportional to $\alpha^{2}$. To be consistent with our perturbative approach, we should expand $t^{-a}$ in powers of $a$. That is, the solution can be expressed
\begin{eqnarray}
    \rho(\vec{p},\vec{p}\,',t)\sim -a\,{\rm ln}\,t+\cdots
\end{eqnarray}
This ln\,$t$ behavior is quite distinctly different from the exponential dependence in the Markovian case in the high temperature limit. 

To incorporate the temperature dependence in this low temperature case, we examine the master equation in Eq.~(\ref{lowmaster}) with the $1/\beta^{2}$ term.
\begin{eqnarray}
    \frac{\partial}{\partial t}\,\rho(\vec{p},\vec{p}\,',t)=\cdots-a\left(\frac{1}{t}+\frac{\pi^{2}\,t}{3\,\beta^{2}}\right)\,\rho(\vec{p},\vec{p}\,',t)
\end{eqnarray}
The solution can be written as
\begin{eqnarray}\label{lowTsol}
    \rho(\vec{p},\vec{p}\,',t)\sim -a\,{\rm ln}\,t-\frac{a\,\pi^{2}}{6\,\beta^{2}}t^{2}+\cdots
\end{eqnarray}
Therefore, the temperature dependent part in this low temperature case is proportional to $t^{2}$ 
{and also proportional $1/\beta^{2}$ (or $T^{2}$).}

{Here, we would like to emphasize that the master equation obtained in Eq.~(\ref{mastereqn2}) (or Eq.~(\ref{FPeqn})) is valid for graviton environments at all  temperatures. At high temperatures with $1/\beta$ being the largest scale, the noise kernel is proportional to a Dirac delta function, one reaches the Markovian limit. In this case, the decoherence of the reduced density matrix elements is exponential with decoherent time scale given by Eq.~(\ref{dectime}), consistent with the results in \cite{AHGraDec} and \cite{Blencowe}. For other temperatures, the dynamics is non-Markovian which can also be analyzed through the master equation in Eq.~(\ref{mastereqn2}). As exemplified by the low temperature limit,  the off-diagonal elements of the reduced density matrix decrease in time  logarithmically for the zero temperature part and quadratically in time for the temperature-dependent part, which is distinctly different from the Markovian case.
}

\section{Discussions} 

\subsection{Three levels of inquiry}

Studies of massive bodies in a gravitational field went as far back as Galileo and Newton, crystallized in foundational principles such as the equivalence principle between inertial and gravitational masses, and the Law of Universal Attraction. Both are in the context of  classical massive objects in a classical gravitational field.  Fast forward to the times of Minkowski and Einstein, we were introduced to a geometric description of the gravitational field, and studies of the weak perturbations from a background spacetime ushers in gravitational wave astronomy and physics. After the advent of quantum mechanics, especially after the triumphant success of quantum electrodynamics in the early 50s, it was natural for theorists to attempt constructing quantum theories of gravity. Quantizing the linearized gravitational perturbations off of the Minkowski metric was easy (compared to the efforts for nonperturbative quantum gravity which also began in the 50s), the short wavelength sector of which can be described by gravitons which is assuring because this spin 2 particle behaves in many ways  like its spin one cousin, the photon.  With this we reach the second level of inquiries,   namely, classical massive systems in a (quantized weak gravitational, or) graviton field.   

Now, more than half a century later, three developments have pushed this inquiry to a higher level still.  They are, the detection of gravitational waves (GW) in 2015, high precision atomic-molecular-optical (AMO) experiments, and advances in quantum information sciences (QI) since the mid-90s\footnote{These advances were interdependent: e.g., advanced quantum optical measurement techniques were instrumental to raising the sensitivity of LIGO detectors. AMO experimental schemes, from ion trap to superconducting qubits to circuit QED, played an important role in realizing quantum information processing schemes.}.  Detection of GWs operates at the first level, that of classical massive systems in a classical gravitational field. Experiments in AMO and QI involving atoms and photons are mostly working with quantum systems in a classical (e.g., laser in a coherent state) or quantum EM field (e.g., cavity QED).   Aided by these rapid advances in observations and experiments one cant help beginning to ponder on the feasibility of detecting gravitons, as a direct evidence of the quantum nature of gravity. Dyson's eloquent essay we mentioned earlier was an effective trigger in that direction. 
Studying how a quantum field, including its fluctuations  (graviton noise),  act  on a classical gravitational systems can offer indirect but distinct evidences of the quantum nature of (perturbative) gravity. See \cite{PWZ,Kanno,ChoHu22} for an example of this second level of inquiry. QI issues such as entanglement between massive objects in a gravitational field opens up the third level of inquiry, where both the massive system and the gravitational field require a quantum treatment.  
Whether the observation of entanglement between quantum massive objects mediated by the Newtonian force would be sufficient to ascertain the quantum nature of  gravity is a currently hot topic of engaging debates. 

As described in the Introduction, in addition to general relativity and quantum field theory for the description of massive systems in a quantized gravitational field, we also need the theory of open quantum systems to describe how  statistical and stochastic properties of the quantum  field environment affect the massive system.  
This paper strives to provide the necessary foundational structure for the inquiry of the above sets of issues. At the second level, we have derived a Langevin equation for the geodesic separation between two {\it classical} masses interacting with a graviton field.  At the third level we have derived a quantum master equation for the reduced density matrix of a quantum massive system interacting with a graviton field and the corresponding FP equation for its associated Wigner distribution function. The challenge is in reaching the non-Markovian regime which are useful for the study of gravitational decoherence and entanglement at low temperatures.

\subsection{Main Features} 

Within this grand structure we summarize the main features of our findings  below.

1. {\it What constitutes a non-Markovian regime? }  We use three parameters,{ $\Lambda$ the cutoff scale of gravitons, $\lambda$ the cutoff scale of the quantum particle, and $1/\beta$ or $T$ the temperature,} to mark the specific ranges and identify the Markovian limit as at a sufficiently high temperature {in which $1/\beta\gg\Lambda\gg\lambda$. In this limit our master equation in Eq.~(\ref{mastereqn2}) or Eq.~(\ref{FPeqn}) are consistent} with the existing Markovian master equations (the ABH master equation).  Anywhere else in this parameter space constitutes the non-Markovian regimes {of which the dynamics can be analyzed by our master equation. As an example, the non-Markovian behaviors at low temperature of the reduced density matrix is given by Eq.~(\ref{lowTsol}).}

2. \textit{Graviton field}: The main characters of the graviton field are the noise and dissipation kernels {as given in Eqs.~(\ref{noise}) and (\ref{dissipation}), respectively}. The noise kernel plays a key role in the analysis of  gravitational decoherence. Its behavior at low  temperatures {(see Eq.~(\ref{lowTnoise}))} is what signifies non-Markovianity and is of special interest in our investigation of a non-Markovian master equation.   The dissipation kernel is independent of temperature and will play a role in gravitational radiation reaction and self force issues.  

3. \textit{Langevin equation for classical massive systems}: In Section II, we   worked out the influence action due to gravitons on a classical massive system.  From there we derived a Langevin equation {(Eq.~(\ref{Langeqn}))}, the solutions of which will enable one to see the effects of graviton field on the geodesic motion of particles.  The  effects of graviton noise on the classical geodesics and their congruences were analyzed in \cite{PWZ,Kanno} and by the present authors \cite{ChoHu22,ChoHu23}.  {Together with the dissipation term, Eq.~(\ref{Langeqn}) is in the form of a MST-QW like equation which can account for the gravitational radiation reaction and self force phenomena as mentioned above.}  

4.  \textit{non-Markovian master equation for quantum massive systems}:  Despite the apparent complexity, the physical meaning of all the terms of, e.g., (\ref{master1}), can be clearly described: The non-trivial dynamics  reside mainly on the interaction terms coming from the noise kernel (the $N_{4}(t)$ term) and the dissipation kernel (the $D_{4}(t)$ term). The other terms are renormalization contributions to the ``free action" of $\vec{z}$ and $\vec{z}\,'$. 

5. \textit{Gravitational Decoherence in the non-Markovian regime}{: At high temperatures (with $1/\beta$ being the largest scale), the time evolution of the off-diagonal elements of the reduced density matrix is an exponential decay with the decoherence constant proportional to the temperature $T$. However, in the low temperature non-Markovian case (see Eq.~(\ref{lowTsol}), the time dependence is logarithmic for the zero temperature part and quadratic for the finite temperature part. Also, the temperature dependent term is proportional to $1/\beta^2$, as contrasted to $1/\beta$ in the high temperature Markovian case. It would be interesting to explore the time evolution of the reduced density matrix element in other temperature regimes by analyzing our master equation in Eq.~(\ref{mastereqn2}) or Eq.~(\ref{FPeqn}).
}

\subsection{Related Problems}

\subsubsection{Gravitational Radiation Reaction and Self Force: Onward to curved spacetimes}

A related problem of importance is gravitational radiation reaction and self force. A similar methodology was used by Galley, Hu and Lin (GLH) \cite{GHL06} to derive the MST-QW equation (referred to in earlier sections) with contributions from the graviton noise.  These authors studied the motion of a small point
mass with stress tensor $T_{\mu\nu}$ moving through a quantized linear weak metric perturbation field $h_{\mu\nu}$ in a curved vacuum background spacetime. The  particle-field dynamics is given in their (GLH) Eq.~(4.1).  The quantum
field fluctuations manifesting as classical stochastic forces act  on the particle, causing a deviation $\tilde z(\tau)$  to its mean value $\bar z(\tau)$ in its trajectory $z(\tau) = \bar z (\tau) + \tilde z(\tau)$, where $\tau$ is the proper time. The semiclassical dynamics for the mean trajectory $\bar z(\tau)$ is given in their Eq.~(4.17) and the stochastic dynamics for the deviations  $\tilde z(\tau)$  from the mean trajectory  is given in their Eq.~(4.24). We bring up this earlier work for two reasons. On a  broader scope,  GHL produced a semiclassical Langevin equation for the motion of a massive particle in curved spacetime which is a generalization of the classical MST-QW equation for the mass's motion including the self force. The GHL equation is semiclassical stochastic in nature as it  includes the backreaction of a quantum field and its fluctuations manifesting as (graviton) noise. Compared with the present work which produces a quantum master equation for the reduced density matrix and a FP equation for the reduced Wigner distribution function of a quantum massive system. The former self force problem can be upgraded to include the stochastic dynamics of {\it quantum massive systems} with the methods of the present work while the present work can be generalized to treating the quantum linear perturbations  of a  {\it general curved spacetime} with the benefit of GHL's earlier study.   

On a narrower scope, statements in GHL related to the dissipation kernel, where the self force originates,  can provide a check on the consistency with our present findings, specifically, by examining the flat space limit of Eq.~(4.24) in GLH. This turns out to be a simple equation (4.25), evident since the tail term vanishes identically in Minkowski spacetime. What we are interested in is the role of the cutoff function on the field modes.   In the paragraph after GLH's Eq.~(4.25)  it is stated, ``At $O(\Lambda^0)$ there is no dissipation term appearing here, which implies that the two-point function of $\dot {\tilde z}$ and $\tilde z$ could grow unbounded in time in the strict point-particle limit $\Lambda \rightarrow \infty$. However, if $\Lambda$ is large but finite then a simple scaling argument implies that dissipation effects from the neglected $O(\Lambda^{-1})$ terms could begin to appear on a time scale of the order of $\Lambda$. Dissipation from higher-order terms arising from the nonlinearities of the full metric perturbation
field equations might begin to appear on a time scale of the order of 
$ 1/m_0$."  Leaving aside the nonlinearities in the metric, since we only treated linear perturbations in this paper, the behavior of $\Lambda$ in our expression for the dissipation kernel  $D_4$ is consistent with this remark from GLH. Once we generalize to consider quantum massive systems in a curved spacetime with quantum gravitational field, gravitational self force will need to be considered. 


\subsubsection{Comparison with quantum Brownian motion (QBM): non-Markovianity and Nonlinearity}
 

In the study of the stochastic dynamics of classical systems Brownian motion is considered the canon, so is quantum Brownian motion (QBM) for quantum systems. The model for a system described by a harmonic oscillator interacting with an environment described by a set of harmonic oscillators, if they are coupled linearly,  thanks to its Gaussian nature, admits an exact quantum master equation \cite{HPZ92} (or the corresponding Fokker-Planck-Wigner equation \cite{HalYu96}) which is valid for all spectral properties of the bath and at all temperatures. It is well known that at moderate to high temperatures thermal noise would decohere most quantum systems rapidly \cite{Zurek}.  This makes the low temperature regime of special interest  because it may permit a longer decoherence time which makes quantum information processing more amenable to practical applications. Here, just as for non-Ohmic baths, one enters the non-Markovian regime. 

Quantitative studies of decoherence based on the exact quantum master equation have been carried out since the early 90s. A good pedagogical example given in \cite{PHZ93} is useful for our comparison discussion here. (For the background, see, e.g.,  the short tutorial \cite{HCH24} on environment-induced decoherence and gravitational decoherence). 
Decoherence originates from the noise in the environment   described by the noise kernel which appears in the quantum diffusion terms in either the master or the FP equation. The form of the master or the FP equation  often found in textbooks contains only one diffusion term (call it `normal' diffusion), and that is because they are derived under high temperature Markovian conditions.  In truth there is another diffusion term,  call it  `anomalous' diffusion. They are proportional to the temperature and the inverse temperature respectively. Thus at low temperatures it is the anomalous diffusion term which  governs the dynamics of decoherence.  This is where the novel non-Markovian feature shows up in environment-induced decoherence.

What we want to do is to understand the differences between decoherence described by the QBM and decoherence by a gravitational field (gravdec) from a quantum master equation such as what is derived here,   Eq.~(\ref{mastereqn2}). The environments  in both cases are pretty much the same, one can use a scalar field for the former \cite{UnrZur,HM94} and a minimally coupled massless scalar field for the latter. The system for QBM is a harmonic oscillator which is often used to describe  the internal (electronic) degrees of freedom of a (harmonic) atom, while the system for gravdec is often a quantum massive object, or the external (mechanical) degree of freedom of a composite particle, such as the center of mass of an atom. The key distinction as far as decoherence is concerned lies in the form of the interaction.  

We have mentioned one key factor  earlier, namely, while quantum systems described by QBM interacting bilinearly in the coordinate variables $(x, q)$  with its environment will decohere most readily in the configuration basis,  gravdec happens in the energy basis because gravitational interactions are via an object's  mass / energy. Another noticeable difference with the linear QBM system is that the interaction term between the gravitons and the quantum particle is linear in $h_{\mu\nu}$ (graviton field) but  quadratic in $\vec{z}$ (particle motion), since the stress-energy of a massive object is proportional to  the square of the object's 4 velocity $u^\mu$.   For this reason,  
we must resort to perturbative approaches \footnote{One can write down the influence action for  interactions of the type $f(x)q$ all right,  which includes the present $x^2q$ type.  It is when one tries to derive the evolution operator that, owing to the existence of terms higher than the second order as in Eq. (\ref{actioncg}), one needs to carry out a perturbative expansion, as in Eq. (\ref{Jxp}). Functional perturbative method has been developed \cite{HPZ93} and applied to QBM of  systems interacting with a bath via $f(x)q^k$ type of coupling. For applications of this method to perturbative treatments of nonlinear QBM and quantum optomechanics, see \cite{SinLopSub,But2Mir}.}.  
Here, to leading order in an expansion in powers of the Newton's constant we see that the noise effect and the dissipation effect separate, with the $N_{4}(t)$ term pertaining to the dissipative phenomena and the $D_{4}(t)$ to the decoherence ones.

\subsubsection{Similarities with Quantum Optomechanics (QOM): facilitating experimental tests} 

Either a moving object such as a cantilever modulating the optical wave modes in a cavity  or the  radiation causing the movement of mechanical objects, both cases fall under QOM.  The latter forms the basis of LIGO type of  gravitational wave detection, which uses the displacement of mirrors upon the impingement of gravitational radiation for its detection. The difference between these two cases lies in what is treated as the system and what is regarded as its environment. The coupling is the same, namely, between the radiation pressure and the displacement of a mechanical device of the $Nq$ type, where $N$ is the number of photons and $q$ is the displacement of the mechanical object.  In a Fock space representation   $N = a^+ a$ where $a^+, a$ are the creation and annihilation operators respectively. When they are expressed in the configuration-momentum basis  $x,p$  we can see that the coupling  is of  quadratic order in the system variables $x$.  On this count it is similar to the appearance of quadratic order of $u^\mu u^\nu$ in the stress energy tensor $T^{\mu\nu}$ of the system variable, which interacts  with the linearized gravitational field $h_{\mu\nu}$  via $T^{\mu\nu} h_{\mu\nu}$. (See the remarks in Sec. II of \cite{ChoHu22} or for a simpler 0 dim QOM model, in \cite{XueBle}:  compare their  Eqs (4) and (8), where the energy of the radiation field is shown explicitly.)  In both cases, decoherence will appear in the energy basis, not in the configuration space basis, as  for the $xq$ type of coupling in QBM. Given this analogy with QOM, the origin  and nature of non-Markovian noise in quantum field environments giving rise to a semiclassical Langevin equation, the way the non-Markovian master and FP equations are derived and how non-Markovian decoherence  for gravitational systems is studied in this paper, they should  provide some useful hints to solving the corresponding theoretical problems  at low temperatures in QOM, such as in dealing with   non-Markovian noise \footnote{Some examples of optomechanical resonators which require treatment of non-Markovian noise and dynamics can be found in a recent review  on massive quantum systems \cite{BoseRMP}: (here we quote)
non-Markovian noise spectrum is observed in a clamped systems \cite{GroHam}, is shown for a nanomechanical flexing beam resonator \cite{RBT}.  Sideband cooling of optomechanical systems in non-Markovian environments was  studied \cite{TEP}, and the influence of non-Markovian noise on the optomechanical nonlinearity
was considered in \cite{Qva}, etc.}, in finding parameter ranges for noise suppression beyond the standard quantum limit, in extending  memory-assisted metrology (e.g., \cite{Jin}) to the quantum domain (e.g., \cite{Chin}), to name just a few, all at the frontier of fundamental theoretical and experimental research.


\vskip 30pt

\noindent{\bf Acknowledgment}  We thank Dr. J.-T. Hsiang for interesting discussions on non-Markovianity.    H.-T. Cho is supported in part by the National Science and Technology Council, Taiwan, R.O.C. under the Grant NSTC 113-2112-M-032-007. B.-L. Hu enjoyed the warm hospitality of Prof.~C.-S. Chu of the National Center for Theoretical Sciences at National Tsing Hua University, and Prof.~K.-W. Ng of the Institute of Physics, Academia Sinica, Taiwan, R.O.C. where a good part of this work was done.

\newpage

\appendix

\section{Regularized representation of the delta function and other related functions}\label{regularized}

In Eq.~(\ref{regdelta}), we have shown that the regularized delta function can be expressed as an integral with an ultraviolet cutoff $\Lambda$
\begin{eqnarray}\label{regdel2}
    \delta_{\Lambda}(t)=\frac{1}{\pi}\int_{0}^{\Lambda}dq\,\cos(q\,t).
\end{eqnarray}
Since the delta function can be expressed as the derivative of the step function $\theta(t)$, we have
\begin{eqnarray}
    \frac{d}{dt}\theta_{\Lambda}(t)=\frac{1}{\pi}\int_{0}^{\Lambda}dq\,\cos(qt)
    \Rightarrow\theta_{\Lambda}(t)=\frac{1}{\pi}\int_{0}^{\Lambda}dq\,\left(\frac{1}{q}\right)\sin(qt)+C
\end{eqnarray}
where $C$ is the integration constant. To determine $C$, we note that $\theta_{\Lambda}(\infty)=1$, and 
\begin{eqnarray}
    \theta_{\Lambda}(\infty)&=&\lim_{t\rightarrow\infty}\frac{1}{\pi}\int_{0}^{\Lambda\,t}dq\,\left(\frac{1}{q}\right)\sin q+C\nonumber\\
    &=&\frac{1}{\pi}\int_{0}^{\infty}dq\,\left(\frac{1}{q}\right)\sin q+C\nonumber\\
    &=&\frac{1}{2}+C
\end{eqnarray}
Hence, $C=1/2$, and we have
\begin{eqnarray}\label{step}
    \theta_{\Lambda}(t)&=&\frac{1}{\pi}\int_{0}^{\Lambda}dq\,\left(\frac{1}{q}\right)\sin(q\,t)+\frac{1}{2}
\end{eqnarray}
Note that in this representation $\theta_{\Lambda}(0)=1/2$.
From Eq.~(\ref{regdel2}), we have
\begin{eqnarray}\label{regdel3}
    \delta_{\Lambda}(t)&=&\frac{1}{\pi}\int_{0}^{\Lambda}dq\,\cos(q\,t)
    =\frac{\sin(\Lambda\,t)}{\pi\,t}
\end{eqnarray}
where $\delta_{\Lambda}(0)=\Lambda/\pi$. Similarly, we have the derivatives
\begin{eqnarray}
    \dot{\delta}_{\Lambda}(t)&=&-\frac{1}{\pi}\int_{0}^{\Lambda}dq\,q\,\sin(q\,t)=\frac{1}{\pi\,t^{2}}\left[\Lambda\, t\cos(\Lambda\,t)-\sin(\Lambda\,t)\right]\\
    \ddot{\delta}_{\Lambda}(t)&=&-\frac{1}{\pi}\int_{0}^{\Lambda}dq\,q^2\,\cos(q\,t)=\frac{1}{\pi\,t^{3}}\left[-2\Lambda\, t\cos(\Lambda\,t)+(2-\Lambda^2 t^2)\sin(\Lambda\,t)\right]
\end{eqnarray}
where $\dot{\delta}(t)|_{t=0}=0$ and $\ddot{\delta}(t)|_{t=0}=-\Lambda^{3}/3\pi$.

With these expressions for the regularized step function, the delta function and its derivatives, one can evaluate the integrals that are introduced in \cite{HH22} that are also used in this paper. First, the integral
\begin{eqnarray}
    I_{0}(t)=\int_{0}^{t}ds\,\delta(t-s)f(s)
\end{eqnarray}
with a well behaved function $f(s)$. Using the regularized integral representation of $\delta_{\Lambda}(t-s)$, $I_{0}(t)$ becomes
\begin{eqnarray}
    I_{0}(t)=\frac{1}{\pi}\int_{0}^{\Lambda}dq\int_{0}^{t}ds\,\cos[q(t-s)]\,f(s)
\end{eqnarray}
In the vicinity of $s=t$, we Taylor-expand $f(s)=f(t)+\dot{f}(t)(s-t)+\cdots$. Then, the integration over $s$ can be evaluated to give
\begin{eqnarray}\label{I0Lambda}
    I_{0}(t)&=&\frac{1}{\pi}f(t)\int_{0}^{\Lambda}dq\int_{0}^{t}ds\,\cos[q(t-s)]+\frac{1}{\pi}\dot{f}(t)\int_{0}^{\Lambda}dq\int_{0}^{t}ds\,(s-t)\cos[q(t-s)]+\cdots\nonumber\\
    &=&\frac{1}{\pi}f(t)\int_{0}^{\Lambda}dq\left(\frac{1}{q}\right)\sin(q\,t)+\frac{1}{\pi}\dot{f}(t)\left(\frac{1}{\Lambda}\right)[1-\cos(\Lambda\,t)]+\cdots
\end{eqnarray}
Note that the second and the subsequent terms vanish as $\Lambda\rightarrow\infty$. In this limit, the first term converges to a step function as shown in Eq.~(\ref{step}).
\begin{eqnarray}\label{I0}
    I_{0}(t)=f(t)\left(\theta(t)-\frac{1}{2}\right)=\frac{1}{2}f(t)[\theta(t)-\theta(-t)]
\end{eqnarray}
which is the same as the result obtained in \cite{HH22}.

Another integral which has been used in this paper is 
\begin{eqnarray}\label{intI1}
    I_{1}(t)=\int_{0}^{t}ds\, [\partial_{s}\delta(t-s)]\,f(s)
\end{eqnarray}
Again, using the regularized integral representation of the delta function in Eq.~(\ref{regdel3} and Taylor-expanding $f(s)$, $I_{1}(t)$ can be expressed in terms of integrals over $q$.
\begin{eqnarray}
    I_{1}(t)&=&\frac{1}{\pi}f(t)\left(\Lambda-\int_{0}^{\Lambda}dq\,\cos(q\,t)\right)\nonumber\\
    &&\ \ +\frac{1}{\pi}\dot{f}(t)\left(t\int_{0}^{\Lambda}dq\,\cos(q\,t)-\int_{0}^{\Lambda}\frac{\sin(q\,t)}{q}\right)+\cdots\nonumber\\
    &=&f(t)[\delta_{\Lambda}(0)-\delta_{\Lambda}(t)]+\dot{f}(t)\left(t\delta_{\Lambda}(t)-\theta_{\Lambda}(t)+\frac{1}{2}\right)+\cdots
\end{eqnarray}
As $\Lambda\rightarrow\infty$, the terms in the ellipsis go to zero. Hence, in this limit,
\begin{eqnarray}
    I_{1}(t)=\delta(0)f(t)-f(0)\delta(t)-\frac{1}{2}\dot{f}(t)[\theta(t)-\theta(-t)]
\end{eqnarray}
which is consistent with the result in \cite{HH22}.

Lastly, we look at another integral,
\begin{eqnarray}
    I_{2}(t)=\partial_{t}\int_{0}^{t}ds\,\delta(t-s)\,f(s)
\end{eqnarray}
which is the derivative of $I_{0}(t)$. With the regularized form of $I_{0}(t)$ in Eq.~(\ref{I0Lambda}), we have
\begin{eqnarray}
    I_{2}(t)&=&\partial_{t}\left[\frac{1}{\pi}f(t)\int_{0}^{\Lambda}dq\left(\frac{1}{q}\right)\sin(q\,t)+\cdots\right]\nonumber\\
    &=&\frac{1}{\pi}f(t)\int_{0}^{\Lambda}dq\,\cos(q\,t)+\frac{1}{\pi}\dot{f}(t)\int_{0}^{\Lambda}dq\left(\frac{1}{q}\right)\sin(q\,t)+\cdots
\end{eqnarray}
where again the ellipsis represents terms which vanish as $\Lambda\rightarrow\infty$. In this limit, we thus have
\begin{eqnarray}
    I_{2}(t)=f(0)\delta(t)+\frac{1}{2}\dot{f}(t)[\theta(t)-\theta(-t)]
\end{eqnarray}
This result can also be obtained if we differentiate the result of $I_{0}(t)$ in Eq.~(\ref{I0}) directly.

\section{Expressions relevant to the derivation of the Eqs.~(\ref{AJnoise}) and (\ref{AJdis})}\label{onJ0}

From the forms of the classical solutions in Eqs.~(\ref{clasols}) and the identities in Eq.~(\ref{identities}), we can derive the following expressions. The ones with only one $\vec{\Delta}_{cl}$ or $\vec{\Sigma}_{cl}$ are
\begin{eqnarray}
    \Delta_{cl}^{i}(t_{1}){\cal J}_{0}(\vec{\Sigma},\vec{\Delta};t|\vec{\Sigma}_{0},\vec{\Delta}_{0};0)&=&\left[\Delta^{i}+i\left(\frac{t-t_{1}}{m}\right)\frac{\partial}{\partial\Sigma^{i}}\right]{\cal J}_{0}(\vec{\Sigma},\vec{\Delta};t|\vec{\Sigma}_{0},\vec{\Delta}_{0};0)\\
    \Sigma_{cl}^{i}(t_{1}){\cal J}_{0}(\vec{\Sigma},\vec{\Delta};t|\vec{\Sigma}_{0},\vec{\Delta}_{0};0)&=&\left[\Sigma^{i}+i\left(\frac{t-t_{1}}{m}\right)\frac{\partial}{\partial\Delta^{i}}\right]{\cal J}_{0}(\vec{\Sigma},\vec{\Delta};t|\vec{\Sigma}_{0},\vec{\Delta}_{0};0)
\end{eqnarray}
The ones with two $\vec{\Delta}_{cl}$ or $\vec{\Sigma}_{cl}$ are
\begin{eqnarray}
    &&\Sigma_{cl}^{i}(t_{1})\Delta_{cl}^{j}(t_{1}){\cal J}_{0}(\vec{\Sigma},\vec{\Delta};t|\vec{\Sigma}_{0},\vec{\Delta}_{0};0)\nonumber\\
    &=&\Bigg[i\,\frac{(t-t_{1})^{2}}{mt}\delta^{ij}+\Sigma^{i}\Delta^{j}+i\left(\frac{t-t_{1}}{m}\right)\left(\Sigma^{i}\frac{\partial}{\partial\Sigma^{j}}+\Delta^{j}\frac{\partial}{\partial\Delta^{i}}\right)-\left(\frac{t-t_{1}}{m}\right)^{2}\frac{\partial}{\partial\Delta^{i}}\frac{\partial}{\partial\Sigma^{j}}\Bigg]\nonumber\\
    &&\hskip 10pt {\cal J}_{0}(\vec{\Sigma},\vec{\Delta};t|\vec{\Sigma}_{0},\vec{\Delta}_{0};0)
\end{eqnarray}
\begin{eqnarray}
    &&\Delta_{cl}^{i}(t_{1})\Sigma_{cl}^{j}(t_{1}){\cal J}_{0}(\vec{\Sigma},\vec{\Delta};t|\vec{\Sigma}_{0},\vec{\Delta}_{0};0)\nonumber\\
    &=&\Bigg[i\,\frac{(t-t_{1})^{2}}{mt}\delta^{ij}+\Delta^{i}\Sigma^{j}+i\left(\frac{t-t_{1}}{m}\right)\left(\Delta^{i}\frac{\partial}{\partial\Delta^{j}}+\Sigma^{j}\frac{\partial}{\partial\Sigma^{i}}\right)-\left(\frac{t-t_{1}}{m}\right)^{2}\frac{\partial}{\partial\Sigma^{i}}\frac{\partial}{\partial\Delta^{j}}\Bigg]\nonumber\\
    &&\hskip 10pt {\cal J}_{0}(\vec{\Sigma},\vec{\Delta};t|\vec{\Sigma}_{0},\vec{\Delta}_{0};0)
\end{eqnarray}
\begin{eqnarray}
    &&\Delta_{cl}^{i}(t_{1})\Delta_{cl}^{j}(t_{1}){\cal J}_{0}(\vec{\Sigma},\vec{\Delta};t|\vec{\Sigma}_{0},\vec{\Delta}_{0};0)\nonumber\\
    &=&\Bigg[\Delta^{i}\Delta^{j}+i\left(\frac{t-t_{1}}{m}\right)\left(\Delta^{i}\frac{\partial}{\partial\Sigma^{j}}+\Delta^{j}\frac{\partial}{\partial\Sigma^{i}}\right)-\left(\frac{t-t_{1}}{m}\right)^{2}\frac{\partial}{\partial\Sigma^{i}}\frac{\partial}{\partial\Sigma^{j}}\Bigg]
    {\cal J}_{0}(\vec{\Sigma},\vec{\Delta};t|\vec{\Sigma}_{0},\vec{\Delta}_{0};0)\nonumber\\
\end{eqnarray}
\begin{eqnarray}
    &&\Sigma_{cl}^{i}(t_{1})\Sigma_{cl}^{j}(t_{1}){\cal J}_{0}(\vec{\Sigma},\vec{\Delta};t|\vec{\Sigma}_{0},\vec{\Delta}_{0};0)\nonumber\\
    &=&\Bigg[\Sigma^{i}\Sigma^{j}+i\left(\frac{t-t_{1}}{m}\right)\left(\Sigma^{i}\frac{\partial}{\partial\Delta^{j}}+\Sigma^{j}\frac{\partial}{\partial\Delta^{i}}\right)-\left(\frac{t-t_{1}}{m}\right)^{2}\frac{\partial}{\partial\Delta^{i}}\frac{\partial}{\partial\Delta^{j}}\Bigg]
    {\cal J}_{0}(\vec{\Sigma},\vec{\Delta};t|\vec{\Sigma}_{0},\vec{\Delta}_{0};0)\nonumber\\
\end{eqnarray}
\begin{eqnarray}
    &&\Sigma_{cl}^{i}(t_{1})\Delta_{cl}^{j}(t_{2}){\cal J}_{0}(\vec{\Sigma},\vec{\Delta};t|\vec{\Sigma}_{0},\vec{\Delta}_{0};0)\nonumber\\
    &=&\Bigg[i\,\frac{(t-t_{1})(t-t_{2})}{mt}\delta^{ij}+\Sigma^{i}\Delta^{j}+i\left(\frac{t-t_{1}}{m}\right)\Delta^{j}\frac{\partial}{\partial\Delta^{i}}+i\left(\frac{t-t_{2}}{m}\right)\Sigma^{i}\frac{\partial}{\partial\Sigma^{j}}+\nonumber\\
    &&\hskip 30pt-\frac{(t-t_{1})(t-t_{2})}{m^2}\frac{\partial}{\partial\Delta^{i}}\frac{\partial}{\partial\Sigma^{j}}\Bigg]
    {\cal J}_{0}(\vec{\Sigma},\vec{\Delta};t|\vec{\Sigma}_{0},\vec{\Delta}_{0};0)
\end{eqnarray}
\begin{eqnarray}
    &&\Delta_{cl}^{i}(t_{1})\Sigma_{cl}^{j}(t_{2}){\cal J}_{0}(\vec{\Sigma},\vec{\Delta};t|\vec{\Sigma}_{0},\vec{\Delta}_{0};0)\nonumber\\
    &=&\Bigg[i\,\frac{(t-t_{1})(t-t_{2})}{mt}\delta^{ij}+\Delta^{i}\Sigma^{j}+i\left(\frac{t-t_{1}}{m}\right)\Sigma^{j}\frac{\partial}{\partial\Sigma^{i}}+i\left(\frac{t-t_{2}}{m}\right)\Delta^{i}\frac{\partial}{\partial\Delta^{j}}\nonumber\\
    &&\hskip 30pt-\frac{(t-t_{1})(t-t_{2})}{m^2}\frac{\partial}{\partial\Sigma^{i}}\frac{\partial}{\partial\Delta^{j}}\Bigg]
    {\cal J}_{0}(\vec{\Sigma},\vec{\Delta};t|\vec{\Sigma}_{0},\vec{\Delta}_{0};0)
\end{eqnarray}
\begin{eqnarray}
    &&\Delta_{cl}^{i}(t_{1})\Delta_{cl}^{j}(t_{2}){\cal J}_{0}(\vec{\Sigma},\vec{\Delta};t|\vec{\Sigma}_{0},\vec{\Delta}_{0};0)\nonumber\\
    &=&\Bigg[\Delta^{i}\Delta^{j}+i\left(\frac{t-t_{1}}{m}\right)\Delta^{j}\frac{\partial}{\partial\Sigma^{i}}+i\left(\frac{t-t_{2}}{m}\right)\Delta^{i}\frac{\partial}{\partial\Sigma^{j}}
    -\frac{(t-t_{1})(t-t_{2})}{m^2}\frac{\partial}{\partial\Sigma^{i}}\frac{\partial}{\partial\Sigma^{j}}\Bigg]\nonumber\\
    &&\hskip 20pt 
    {\cal J}_{0}(\vec{\Sigma},\vec{\Delta};t|\vec{\Sigma}_{0},\vec{\Delta}_{0};0)\nonumber\\
\end{eqnarray}
\begin{eqnarray}
    &&\Sigma_{cl}^{i}(t_{1})\Sigma_{cl}^{j}(t_{2}){\cal J}_{0}(\vec{\Sigma},\vec{\Delta};t|\vec{\Sigma}_{0},\vec{\Delta}_{0};0)\nonumber\\
    &=&\Bigg[\Sigma^{i}\Sigma^{j}+i\left(\frac{t-t_{1}}{m}\right)\Sigma^{j}\frac{\partial}{\partial\Delta^{i}}+i\left(\frac{t-t_{2}}{m}\right)\Sigma^{i}\frac{\partial}{\partial\Delta^{j}}-\frac{(t-t_{1})(t-t_{2})}{m^2}\frac{\partial}{\partial\Delta^{i}}\frac{\partial}{\partial\Delta^{j}}\Bigg]\nonumber\\
    &&\hskip 20pt 
    {\cal J}_{0}(\vec{\Sigma},\vec{\Delta};t|\vec{\Sigma}_{0},\vec{\Delta}_{0};0)\nonumber\\
\end{eqnarray}
Lastly, for the ones with four $\vec{\Delta}_{cl}$ and $\vec{\Sigma}_{cl}$ are the following.
\begin{eqnarray}
     &&\Delta_{cl}^{i}(t_{1})\Sigma_{cl}^{j}(t_{1})\Sigma_{cl}^{k}(t_{2})\Sigma_{cl}^{l}(t_{2}){\cal J}_{0}(\vec{\Sigma},\vec{\Delta};t|\vec{\Sigma}_{0},\vec{\Delta}_{0};0)\nonumber\\
     &=&\Bigg[\Delta^{i}\Sigma^{j}\Sigma^{k}\Sigma^{l}+i\frac{(t-t_{1})^2}{m\,t}\delta^{ij}\Sigma^{k}\Sigma^{l}+i\frac{(t-t_{1})(t-t_{2})}{m\,t}\left(\delta^{ik}\Sigma^{j}\Sigma^{l}+\delta^{il}\Sigma^{j}\Sigma^{k}\right)\nonumber\\
     &&\ \ +i\frac{(t-t_{1})}{m}\Sigma^{k}\Sigma^{l}\left(\Sigma^{j}\frac{\partial}{\partial\Sigma^{i}}+\Delta^{i}\frac{\partial}{\partial\Delta^{j}}\right)+i\frac{(t-t_{2})}{m}\Delta^{i}\Sigma^{j}\left(\Sigma^{k}\frac{\partial}{\partial\Delta^{l}}+\Sigma^{l}\frac{\partial}{\partial\Delta^{k}}\right)\nonumber\\
     &&\ \ -\frac{(t-t_{1})^{2}(t-t_{2})}{m^2\,t}\left(\delta^{ij}\Sigma^{k}\frac{\partial}{\partial\Delta^{l}}+\delta^{ij}\Sigma^{l}\frac{\partial}{\partial\Delta^{k}}+\delta^{il}\Sigma^{k}\frac{\partial}{\partial\Delta^{j}}+\delta^{ik}\Sigma^{l}\frac{\partial}{\partial\Delta^{j}}\right)\nonumber\\
     &&\ \ -\frac{(t-t_{1})(t-t_{2})^{2}}{m^{2}t}\left(\delta^{il}\Sigma^{j}\frac{\partial}{\partial\Delta^{k}}+\delta^{ik}\Sigma^{j}\frac{\partial}{\partial\Delta^{l}}\right)\nonumber\\
     &&\ \ -\frac{(t-t_{1})(t-t_{2})}{m^{2}}\bigg(\Sigma^{j}\Sigma^{k}\frac{\partial}{\partial\Sigma^{i}}\frac{\partial}{\partial\Delta^{l}}+\Sigma^{j}\Sigma^{l}\frac{\partial}{\partial\Sigma^{i}}\frac{\partial}{\partial\Delta^{k}}\nonumber\\
     &&\hskip 120pt
     +\Delta^{i}\Sigma^{k}\frac{\partial}{\partial\Delta^{j}}\frac{\partial}{\partial\Delta^{l}}+\Delta^{i}\Sigma^{l}\frac{\partial}{\partial\Delta^{j}}\frac{\partial}{\partial\Delta^{k}}\bigg)\nonumber\\
     &&\ \ -\frac{(t-t_{1})^2}{m^2}\Sigma^{k}\Sigma^{l}\frac{\partial}{\partial\Sigma^{i}}\frac{\partial}{\partial\Delta^{j}}-\frac{(t-t_{2})^{2}}{m^{2}}\Delta^{i}\Sigma^{j}\frac{\partial}{\partial\Delta^{k}}\frac{\partial}{\partial\Delta^{l}}\nonumber\\
     &&\ \ -i\frac{(t-t_{1})^{2}(t-t_{2})^{2}}{m^3 t}\left(\delta^{ik}\frac{\partial}{\partial\Delta^{j}}\frac{\partial}{\partial\Delta^{l}}+\delta^{il}\frac{\partial}{\partial\Delta^{j}}\frac{\partial}{\partial\Delta^{k}}+\delta^{ij}\frac{\partial}{\partial\Delta^{k}}\frac{\partial}{\partial\Delta^{l}}\right)\nonumber\\
     &&\ \ -i\frac{(t-t_{1})^{2}(t-t_{2})}{m^3}\left(\Sigma^{k}\frac{\partial}{\partial\Sigma^{i}}\frac{\partial}{\partial\Delta^{j}}\frac{\partial}{\partial\Delta^{l}}+\Sigma^{l}\frac{\partial}{\partial\Sigma^{i}}\frac{\partial}{\partial\Delta^{j}}\frac{\partial}{\partial\Delta^{k}}\right)\nonumber\\
     &&\ \ -i\frac{(t-t_{1})(t-t_{2})^{2}}{m^3}\left(\Delta^{i}\frac{\partial}{\partial\Delta^{j}}\frac{\partial}{\partial\Delta^{k}}\frac{\partial}{\partial\Delta^{l}}+\Sigma^{j}\frac{\partial}{\partial\Sigma^{i}}\frac{\partial}{\partial\Delta^{k}}\frac{\partial}{\partial\Delta^{l}}\right)\nonumber\\
     &&\ \ +\frac{(t-t_{1})^2(t-t_{2})^{2}}{m^{4}}\frac{\partial}{\partial\Sigma^{i}}\frac{\partial}{\partial\Delta^{j}}\frac{\partial}{\partial\Delta^{k}}\frac{\partial}{\partial\Delta^{l}}\Bigg]{\cal J}_{0}(\vec{\Sigma},\vec{\Delta};t|\vec{\Sigma}_{0},\vec{\Delta}_{0};0)
\end{eqnarray}
\begin{eqnarray}
     &&\Sigma_{cl}^{i}(t_{1})\Delta_{cl}^{j}(t_{1})\Delta_{cl}^{k}(t_{2})\Delta_{cl}^{l}(t_{2}){\cal J}_{0}(\vec{\Sigma},\vec{\Delta};t|\vec{\Sigma}_{0},\vec{\Delta}_{0};0)\nonumber\\
     &=&\Bigg[\Sigma^{i}\Delta^{j}\Delta^{k}\Delta^{l}+i\frac{(t-t_{1})^2}{m\,t}\delta^{ij}\Delta^{k}\Delta^{l}+i\frac{(t-t_{1})(t-t_{2})}{m\,t}\left(\delta^{ik}\Delta^{j}\Delta^{l}+\delta^{il}\Delta^{j}\Delta^{k}\right)\nonumber\\
     &&\ \ +i\frac{(t-t_{1})}{m}\Delta^{k}\Delta^{l}\left(\Sigma^{i}\frac{\partial}{\partial\Sigma^{j}}+\Delta^{j}\frac{\partial}{\partial\Delta^{i}}\right)+i\frac{(t-t_{2})}{m}\Sigma^{i}\Delta^{j}\left(\Delta^{k}\frac{\partial}{\partial\Sigma^{l}}+\Delta^{l}\frac{\partial}{\partial\Sigma^{k}}\right)\nonumber\\
     &&\ \ -\frac{(t-t_{1})^{2}(t-t_{2})}{m^2\,t}\left(\delta^{ij}\Delta^{k}\frac{\partial}{\partial\Sigma^{l}}+\delta^{ij}\Delta^{l}\frac{\partial}{\partial\Sigma^{k}}+\delta^{il}\Delta^{k}\frac{\partial}{\partial\Sigma^{j}}+\delta^{ik}\Delta^{l}\frac{\partial}{\partial\Sigma^{j}}\right)\nonumber\\
     &&\ \ -\frac{(t-t_{1})(t-t_{2})^{2}}{m^{2}t}\left(\delta^{il}\Delta^{j}\frac{\partial}{\partial\Sigma^{k}}+\delta^{ik}\Delta^{j}\frac{\partial}{\partial\Sigma^{l}}\right)\nonumber\\
     &&\ \ -\frac{(t-t_{1})(t-t_{2})}{m^{2}}\bigg(\Delta^{j}\Delta^{k}\frac{\partial}{\partial\Delta^{i}}\frac{\partial}{\partial\Sigma^{l}}+\Delta^{j}\Delta^{l}\frac{\partial}{\partial\Delta^{i}}\frac{\partial}{\partial\Sigma^{k}}\nonumber\\
     &&\hskip 120pt
     +\Sigma^{i}\Delta^{l}\frac{\partial}{\partial\Sigma^{j}}\frac{\partial}{\partial\Sigma^{k}}+\Sigma^{i}\Delta^{k}\frac{\partial}{\partial\Sigma^{j}}\frac{\partial}{\partial\Sigma^{l}}\bigg)\nonumber\\
     &&\ \ -\frac{(t-t_{1})^2}{m^2}\Delta^{k}\Delta^{l}\frac{\partial}{\partial\Delta^{i}}\frac{\partial}{\partial\Sigma^{j}}-\frac{(t-t_{2})^{2}}{m^{2}}\Sigma^{i}\Delta^{j}\frac{\partial}{\partial\Sigma^{k}}\frac{\partial}{\partial\Sigma^{l}}\nonumber\\
     &&\ \ -i\frac{(t-t_{1})^{2}(t-t_{2})^{2}}{m^3 t}\left(\delta^{ik}\frac{\partial}{\partial\Sigma^{j}}\frac{\partial}{\partial\Sigma^{l}}+\delta^{il}\frac{\partial}{\partial\Sigma^{j}}\frac{\partial}{\partial\Sigma^{k}}+\delta^{ij}\frac{\partial}{\partial\Sigma^{k}}\frac{\partial}{\partial\Sigma^{l}}\right)\nonumber\\
     &&\ \ -i\frac{(t-t_{1})^{2}(t-t_{2})}{m^3}\left(\Delta^{k}\frac{\partial}{\partial\Delta^{i}}\frac{\partial}{\partial\Sigma^{j}}\frac{\partial}{\partial\Sigma^{l}}+\Delta^{l}\frac{\partial}{\partial\Delta^{i}}\frac{\partial}{\partial\Sigma^{j}}\frac{\partial}{\partial\Sigma^{k}}\right)\nonumber\\
     &&\ \ -i\frac{(t-t_{1})(t-t_{2})^{2}}{m^3}\left(\Sigma^{i}\frac{\partial}{\partial\Sigma^{j}}\frac{\partial}{\partial\Sigma^{k}}\frac{\partial}{\partial\Sigma^{l}}+\Delta^{j}\frac{\partial}{\partial\Delta^{i}}\frac{\partial}{\partial\Sigma^{k}}\frac{\partial}{\partial\Sigma^{l}}\right)\nonumber\\
     &&\ \ +\frac{(t-t_{1})^2(t-t_{2})^{2}}{m^{4}}\frac{\partial}{\partial\Delta^{i}}\frac{\partial}{\partial\Sigma^{j}}\frac{\partial}{\partial\Sigma^{k}}\frac{\partial}{\partial\Sigma^{l}}\Bigg]{\cal J}_{0}(\vec{\Sigma},\vec{\Delta};t|\vec{\Sigma}_{0},\vec{\Delta}_{0};0)
\end{eqnarray}
\begin{eqnarray}
     &&\Delta_{cl}^{i}(t_{1})\Sigma_{cl}^{j}(t_{1})\Delta_{cl}^{k}(t_{2})\Sigma_{cl}^{l}(t_{2}){\cal J}_{0}(\vec{\Sigma},\vec{\Delta};t|\vec{\Sigma}_{0},\vec{\Delta}_{0};0)\nonumber\\
     &=&\Bigg[-\frac{(t-t_{1})^2(t-t_{2})^2}{m^2 t^2}\left(\delta^{ij}\delta^{kl}+\delta^{il}\delta^{jk}\right)+\Delta^{i}\Sigma^{j}\Delta^{k}\Sigma^{l}+i\frac{(t-t_{1})^2}{m\,t}\delta^{ij}\Delta^{k}\Sigma^{l}\nonumber\\
     &&\ \ +i\frac{(t-t_{2})^2}{m\,t}\delta^{kl}\Delta^{i}\Sigma^{j}
     +i\frac{(t-t_{1})(t-t_{2})}{m\,t}\left(\delta^{il}\Sigma^{j}\Delta^{k}+\delta^{jk}\Delta^{i}\Sigma^{l}\right)\nonumber\\
     &&\ \ +i\frac{(t-t_{1})}{m}\Delta^{k}\Sigma^{l}\left(\Sigma^{j}\frac{\partial}{\partial\Sigma^{i}}+\Delta^{i}\frac{\partial}{\partial\Delta^{j}}\right)+i\frac{(t-t_{2})}{m}\Delta^{i}\Sigma^{j}\left(\Sigma^{l}\frac{\partial}{\partial\Sigma^{k}}+\Delta^{k}\frac{\partial}{\partial\Delta^{l}}\right)\nonumber\\
     &&\ \ -\frac{(t-t_{1})^{2}(t-t_{2})}{m^2\,t}\left(\delta^{ij}\Delta^{k}\frac{\partial}{\partial\Delta^{l}}+\delta^{ij}\Sigma^{l}\frac{\partial}{\partial\Sigma^{k}}+\delta^{il}\Delta^{k}\frac{\partial}{\partial\Delta^{j}}+\delta^{jk}\Sigma^{l}\frac{\partial}{\partial\Sigma^{i}}\right)\nonumber\\
     &&\ \ -\frac{(t-t_{1})(t-t_{2})^{2}}{m^{2}t}\left(\delta^{kl}\Delta^{i}\frac{\partial}{\partial\Delta^{j}}+\delta^{kl}\Sigma^{j}\frac{\partial}{\partial\Sigma^{i}}+\delta^{il}\Sigma^{j}\frac{\partial}{\partial\Sigma^{k}}+\delta^{jk}\Delta^{i}\frac{\partial}{\partial\Delta^{l}}\right)\nonumber\\
     &&\ \ -\frac{(t-t_{1})(t-t_{2})}{m^{2}}\bigg(\Sigma^{j}\Sigma^{l}\frac{\partial}{\partial\Sigma^{i}}\frac{\partial}{\partial\Sigma^{k}}+\Delta^{i}\Sigma^{l}\frac{\partial}{\partial\Sigma^{k}}\frac{\partial}{\partial\Delta^{j}}\nonumber\\
     &&\hskip 120pt
     +\Sigma^{j}\Delta^{k}\frac{\partial}{\partial\Sigma^{i}}\frac{\partial}{\partial\Delta^{l}}+\Delta^{i}\Delta^{k}\frac{\partial}{\partial\Delta^{j}}\frac{\partial}{\partial\Delta^{l}}\bigg)\nonumber\\
     &&\ \ -\frac{(t-t_{1})^2}{m^2}\Delta^{k}\Sigma^{l}\frac{\partial}{\partial\Sigma^{i}}\frac{\partial}{\partial\Delta^{j}}-\frac{(t-t_{2})^{2}}{m^{2}}\Delta^{i}\Sigma^{j}\frac{\partial}{\partial\Sigma^{k}}\frac{\partial}{\partial\Delta^{l}}\nonumber\\
     &&\ \ -i\frac{(t-t_{1})^{2}(t-t_{2})^{2}}{m^3 t}\left(\delta^{ij}\frac{\partial}{\partial\Sigma^{k}}\frac{\partial}{\partial\Delta^{l}}+\delta^{jk}\frac{\partial}{\partial\Sigma^{i}}\frac{\partial}{\partial\Delta^{l}}+\delta^{il}\frac{\partial}{\partial\Sigma^{k}}\frac{\partial}{\partial\Delta^{j}}+\delta^{kl}\frac{\partial}{\partial\Sigma^{i}}\frac{\partial}{\partial\Delta^{j}}\right)\nonumber\\
     &&\ \ -i\frac{(t-t_{1})^{2}(t-t_{2})}{m^3}\left(\Sigma^{l}\frac{\partial}{\partial\Sigma^{i}}\frac{\partial}{\partial\Sigma^{k}}\frac{\partial}{\partial\Delta^{j}}+\Delta^{k}\frac{\partial}{\partial\Sigma^{i}}\frac{\partial}{\partial\Delta^{j}}\frac{\partial}{\partial\Delta^{l}}\right)\nonumber\\
     &&\ \ -i\frac{(t-t_{1})(t-t_{2})^{2}}{m^3}\left(\Sigma^{j}\frac{\partial}{\partial\Sigma^{i}}\frac{\partial}{\partial\Sigma^{k}}\frac{\partial}{\partial\Delta^{l}}+\Delta^{i}\frac{\partial}{\partial\Sigma^{k}}\frac{\partial}{\partial\Delta^{j}}\frac{\partial}{\partial\Delta^{l}}\right)\nonumber\\
     &&\ \ +\frac{(t-t_{1})^2(t-t_{2})^{2}}{m^{4}}\frac{\partial}{\partial\Sigma^{i}}\frac{\partial}{\partial\Sigma^{k}}\frac{\partial}{\partial\Delta^{j}}\frac{\partial}{\partial\Delta^{l}}\Bigg]{\cal J}_{0}(\vec{\Sigma},\vec{\Delta};t|\vec{\Sigma}_{0},\vec{\Delta}_{0};0)
\end{eqnarray}

\section{Coefficient functions of the evolution operator and the master equation}
In this appendix we consider in some more details the coefficients of the evolution operator as well as those of the master equation.

\subsection{The evolution operator}
The coefficient functions of the evolution operator are given by Eqs.~(\ref{n1}) to (\ref{d4}).  In particular, we shall derive their low temperature and high temperature expansions. Let us start with the coefficient $n_{1}(t)$ in Eq.~(\ref{n1}). We first put in an ultraviolet cutoff $\Lambda$ in the $q$-integral in the kernel $N(t_{1},t_{2})$. 
\begin{eqnarray}
    N(t_{1},t_{2})&=&-\frac{32\pi^4\alpha^2}{15}\int_{0}^{\Lambda}dq\,q\,\cos[q(t_{1}-t_{2})]\left(1+\frac{2}{e^{q\beta}-1}\right)
\end{eqnarray}
Taking the derivatives of the Green function $G(t_{1},t_{2})$  with respect to $t_{1}$ and $t_{2}$ gives
\begin{eqnarray}
    &&\left(\frac{d^{2}}{dt_{1}^{2}}\right)\left(\frac{d^{2}}{dt_{2}^{2}}\right)[(t-t_{1})(t-t_{2})G(t_{1},t_{2})]\nonumber\\
    &=&\frac{4}{t}-8\,\delta_{\lambda}(t_{1}-t_{2})-6\,(t_{1}-t_{2})\,\frac{d}{dt_{1}}\delta_{\lambda}(t_{1}-t_{2})+(t-t_{1})(t-t_{2})\,\frac{d^{2}}{dt_{1}^{2}}\delta_{\lambda}(t_{1}-t_{2})
\end{eqnarray}
where we have introduced a cutoff scale $\lambda$ for the Green function of the quantum particle with the regularized delta function given in Eq.~(\ref{deltalambda}).

The zero temperature  
part of $n_{1}(t)$ can now be expressed as
\begin{eqnarray}
    n_{1}^{(0)}(t)&=&\frac{256\pi^{4}\alpha^{2}}{m^{2}}\int_{0}^{\Lambda}dq\,q\int_{0}^{t}dt_{1}\int_{0}^{t}dt_{2}\,\cos[q(t_{1}-t_{2})]\bigg[-\frac{2}{t^{2}}+\frac{8}{t}\,\delta_{\lambda}(t_{1}-t_{2})\nonumber\\
    &&\hskip 80pt+\frac{6}{t}\,(t_{1}-t_{2})\,\frac{d}{dt_{1}}\delta_{\lambda}(t_{1}-t_{2})-\frac{1}{t}(t-t_{1})(t-t_{2})\,\frac{d^{2}}{dt_{1}^{2}}\delta_{\lambda}(t_{1}-t_{2})\bigg]\nonumber\\
\end{eqnarray}
After doing the integrations, we can expand the result for large $\Lambda$ and $\lambda$ but keeping $\Lambda\gg\lambda$. 
\begin{eqnarray}
    n_{1}^{(0)}(t)&=&\frac{256\pi^{4}\alpha^{2}}{m^{2}}\Bigg[-\frac{4}{t^{2}}\left({\rm ln}(\Lambda\,t)+\gamma\right)+\frac{\lambda}{\pi t}\left(\frac{1}{3}\lambda^{2}t^{2}+16\right)\left({\rm ln}\Lambda-{\rm ln}\lambda\right)\nonumber\\
    &&\hskip 90pt +\frac{\lambda}{t}\left(\frac{1}{12}\lambda^{3}t^{3}-\frac{2}{9\pi}\lambda^{2}t^{2}+7\lambda\, t-\frac{35}{3\pi}\right)+\cdots\Bigg]
\end{eqnarray}
where $\gamma$ is the Euler's constant and the ellipsis represents terms which will vanish as $\Lambda$ and $\lambda$ go to infinity.

For the thermal or finite temperature part,
\begin{eqnarray}\label{n1lowtemp}
    n_{1}^{(\beta)}(t)&=&\frac{256\pi^{4}\alpha^{2}}{m^{2}}\int_{0}^{\Lambda}dq\,q\int_{0}^{t}dt_{1}\int_{0}^{t}dt_{2}\,\cos[q(t_{1}-t_{2})]\left(\frac{2}{e^{q\beta}-1}\right)\bigg[-\frac{2}{t^{2}}+\frac{8}{t}\,\delta_{\lambda}(t_{1}-t_{2})\nonumber\\
    &&\hskip 80pt+\frac{6}{t}\,(t_{1}-t_{2})\,\frac{d}{dt_{1}}\delta_{\lambda}(t_{1}-t_{2})-\frac{1}{t}(t-t_{1})(t-t_{2})\,\frac{d^{2}}{dt_{1}^{2}}\delta_{\lambda}(t_{1}-t_{2})\bigg]\nonumber\\
\end{eqnarray}
As we have discussed in Section II, the $q$-integral here is in fact convergent as given in Eq.~(\ref{thermalnoise}). That is, one can take the limit $\Lambda\rightarrow\infty$. For the low temperature limit, we expand the result in powers of $1/\beta$.
\begin{eqnarray}\label{low}
    \int_{0}^{\infty}dq\,q\,\cos[q(t_{1}-t_{2})]\left(\frac{2}{e^{q\beta}-1}\right)=\frac{\pi^2}{3\beta^2}-\frac{\pi^4 (t_{1}-t_{2})^2}{15\beta^4}+\frac{2\pi^6 (t_{1}-t_{2})^4}{189\beta^6}+\cdots
\end{eqnarray}
Putting this back into Eq.~(\ref{n1lowtemp}), the $t$-integrations can be done readily to give
\begin{eqnarray}
    n_{1}^{(\beta)}(t)=\frac{256\pi^{4}\alpha^{2}}{m^{2}}\left[\frac{\pi\lambda\,t}{3\beta^{2}}+\frac{\pi^{4}t^{2}}{15\beta^{4}}-\frac{4\pi^{6}t^{4}}{2835\beta^{6}}+\cdots\right]
\end{eqnarray}
Together with the zero temperature part, this coefficient function in the low temperature expansion with $\Lambda\gg\lambda\gg 1/t\gg 1/\beta$ is
\begin{eqnarray}
    n_{1}(t)&=&n_{1}^{(0)}(t)+n_{1}^{(\beta)}(t)\nonumber\\
    &=&\frac{256\pi^{4}\alpha^{2}}{m^{2}}\Bigg[-\frac{4}{t^{2}}\left({\rm ln}(\Lambda\,t)+\gamma\right)+\frac{\lambda}{\pi t}\left(\frac{1}{3}\lambda^{2}t^{2}+16\right)\left({\rm ln}\Lambda-{\rm ln}\lambda\right)\nonumber\\
    &&\hskip 50pt +\frac{\lambda}{t}\left(\frac{1}{12}\lambda^{3}t^{3}-\frac{2}{9\pi}\lambda^{2}t^{2}+7\lambda\, t-\frac{35}{3\pi}\right)
    +\frac{\pi\lambda\, t}{3\beta^{2}}+\frac{\pi^{4}t^{2}}{15\beta^{4}}-\frac{4\pi^{6}t^{4}}{2835\beta^{6}}+\cdots\Bigg]\nonumber\\
\end{eqnarray}

When we try to develop a high temperature limit, we shall be interested in the case with $1/\beta\gg\Lambda\gg\lambda\gg 1/t$. As we have discussed in Section II, this would correspond to the Markovian limit in which the leading behavior of the noise kernel is that of a delta function. Since $1/\beta\gg\Lambda$, one cannot develop this expansion by taking the $\Lambda\rightarrow\infty$ limit first as we have done in getting the low temperature expansion in Eq.~(\ref{low}). On the other hand, with $1/\beta\gg\Lambda$, $q\beta$ is always small because the upper limit of the $q$-integration is just $\Lambda$. Hence, one can expand
\begin{eqnarray}
    1+\frac{2}{e^{q\beta}-1}=\frac{2}{q\beta}+\frac{q\beta}{6}+\cdots
\end{eqnarray}
as a power series in $\beta$. Note that the $\beta$ independent terms canceled each other. Finally, in the high temperature limit, with $1/\beta\gg\Lambda\gg\lambda\gg 1/t$, the coefficient function becomes
\begin{eqnarray}
    n_{1}(t)&=&\frac{256\pi^{4}\alpha^{2}}{m^{2}}\int_{0}^{\Lambda}dq\,q\int_{0}^{t}dt_{1}\int_{0}^{t}dt_{2}\,\cos[q(t_{1}-t_{2})]\left(\frac{2}{q\beta}+\frac{q\beta}{6}+\cdots\right)\bigg[-\frac{2}{t^{2}}\nonumber\\
    &&\hskip 20pt+\frac{8}{t}\,\delta_{\lambda}(t_{1}-t_{2})+\frac{6}{t}\,(t_{1}-t_{2})\,\frac{d}{dt_{1}}\delta_{\lambda}(t_{1}-t_{2})-\frac{1}{t}(t-t_{1})(t-t_{2})\,\frac{d^{2}}{dt_{1}^{2}}\delta_{\lambda}(t_{1}-t_{2})\bigg]\nonumber\\
    &=&\frac{256\pi^{4}\alpha^{2}}{m^{2}}\Bigg\{\frac{1}{\beta}\left(\frac{2}{t}\right)\left(\frac{1}{9}\lambda^{3}t^{3}+8\lambda\, t-2\pi\right)\nonumber\\
    &&\hskip 60pt +\beta\bigg[\frac{\Lambda}{t^{2}}\left(\frac{1}{18\pi}\lambda^{3}t^{3}+\frac{8}{3\pi}\lambda\, t-\frac{2}{3}\right)+\frac{\lambda^{3}}{6}\left(\frac{1}{15}\lambda^{2}t^{2}+\frac{20}{3}\right)\nonumber\\
    &&\hskip 90pt +\frac{2}{3\,t^{3}}\sin(\Lambda\, t)\left(1-\frac{3}{\pi}\lambda\, t\cos(\lambda\, t)-\frac{1}{\pi}\sin(\lambda\, t)\right)\Bigg]+\cdots\Bigg\}
\end{eqnarray}

Similar considerations can be applied to the coefficients $n_{2}(t)$ to $n_{4}(t)$. For the low temperature expansions, with $\Lambda\gg\lambda\gg 1/t\gg 1/\beta$, we have
\begin{eqnarray}
    n_{2}(t)&=&\frac{256\pi^{4}\alpha^{2}}{3m}\Bigg\{\left[-\left(\frac{2}{3\pi}\right)\lambda^{3}({\rm ln}\Lambda-{\rm ln}\lambda)+\lambda^{3}\left(-\frac{1}{4}\lambda\,t+\frac{4}{9\pi}\right)-\frac{2}{\pi t^{3}}\sin(\lambda\,t)\right]\nonumber\\
    &&\hskip 80pt +\frac{1}{\beta^{2}}\left[-\frac{2\pi}{3}\lambda+\frac{2\pi}{3\,t}\sin(\lambda\,t)\right]+\frac{1}{\beta^{4}}\left[-\frac{2\pi^{4}t}{15}-\frac{2\pi^{3}t}{15}\sin(\lambda\,t)\right]\nonumber\\
    &&\hskip 100pt +\frac{1}{\beta^{6}}\left[\frac{4\pi^{5}t^{3}}{189}\sin(\lambda\,t)\right]+\cdots\Bigg\}
\end{eqnarray}
\begin{eqnarray}
    n_{3}(t)&=&\frac{256\pi^{4}\alpha^{2}}{3m^{2}}\Bigg\{\left[\frac{\lambda}{\pi}\left(\frac{1}{3}\lambda^{2}t^{2}+16\right)({\rm ln}\Lambda-{\rm ln}\lambda)+\lambda\left(\frac{1}{12}\lambda^{3}t^{3}-\frac{2}{9\pi}\lambda^{2}t^{2}+7\lambda\,t-\frac{35}{3\pi}\right)\right]\nonumber\\
    &&\hskip 60pt +\frac{1}{\beta^{2}}\left(\frac{\pi t}{3}\right)\left(\lambda\,t+2\pi\right)+\frac{1}{\beta^{4}}\left(\frac{2\pi^{4}t^{3}}{45}\right)+\cdots\Bigg\}
\end{eqnarray}
\begin{eqnarray}
    n_{4}(t)&=&-\frac{512\pi^{4}\alpha^{2}}{15m^{4}}\left[\left({\rm ln}(\Lambda\,t)+\gamma\right)+\frac{\pi^{2}t^{2}}{6\beta^{2}}-\frac{\pi^{4}t^{4}}{180\beta^{4}}+\frac{\pi^{6}t^{6}}{2835\beta^{6}}+\cdots\right]
\end{eqnarray}

For the high temperature limit, with $1/\beta\gg\Lambda\gg\lambda\gg 1/t$, we have
\begin{eqnarray}
    n_{2}(t)&=&\frac{256\pi^{4}\alpha^{2}}{3m}\Bigg\{\frac{1}{\beta}\left(-\frac{2}{3}\lambda^{3}t\right)+\beta\bigg[-\frac{1}{9\pi}\Lambda\lambda^{3}-\frac{1}{30}\lambda^{5}t\nonumber\\
    &&\hskip 60pt+\frac{1}{3\pi t^{4}}\sin(\Lambda\,t)\left(\lambda^{2}t^{2}\sin(\lambda\,t)+2\lambda\,t\cos(\lambda\,t)-2\sin(\lambda\,t)\right)\bigg]+\cdots\Bigg\}\nonumber\\
\end{eqnarray}
\begin{eqnarray}
    n_{3}(t)&=&\frac{256\pi^{4}\alpha^{2}}{3m^{2}}\Bigg\{\frac{1}{\beta}\left(\frac{2}{9}\lambda^{3}t^{3}+16\lambda\,t\right)+\beta\bigg[\frac{\Lambda\lambda}{\pi}\left(\frac{1}{18}\lambda^{2}t^{2}+\frac{8}{3}\right)+\frac{\lambda^{3}t}{9}\left(\frac{1}{10}\lambda^{2}t^{2}+10\right)\nonumber\\
    &&\hskip 80pt-\frac{2}{\pi t^{2}}\sin(\Lambda\,t)\left(\lambda\,t\cos(\lambda\,t)+\frac{1}{3}\sin(\lambda\,t)\right)\bigg]+\cdots\Bigg\}
\end{eqnarray}
\begin{eqnarray}
    n_{4}(t)=-\frac{256\pi^{4}\alpha^{2}}{15m^{4}}\left[\frac{1}{\beta}\left(2\pi t\right)+\beta\left(\frac{\Lambda}{3}-\frac{1}{3\,t}\sin(\Lambda\,t)\right)+\cdots\right]
\end{eqnarray}

Next, we look at the coefficients $d_{1}(t)$ to $d_{4}(t)$ which involve the kernel 
\begin{eqnarray}
    D(t_{1},t_{2})=\frac{64\pi^{5}\alpha^{2}}{15}\theta(t_{1}-t_{2})\left[\frac{d}{dt_{1}}\delta_{\Lambda}(t_{1}-t_{2})\right]
\end{eqnarray}
which is expressed as the time derivative of a delta function. Hence, the dissipation part of our effective action is actually Markovian as $\Lambda\rightarrow\infty$. The coefficient function $d_{1}(t)$ in Eq.~(\ref{d1}) can therefore be expressed as
\begin{eqnarray}
    d_{1}(t)=-\frac{512\pi^{5}\alpha^{2}}{3m}\int_{0}^{t}dt_{1}\int_{0}^{t_{1}}dt_{2}\left[\frac{d^{2}}{dt_{1}^{2}}\delta_{\lambda}(t_{1}-t_{2})\right]\left[\frac{d}{dt_{1}}\delta_{\Lambda}(t_{1}-t_{2})\right]
\end{eqnarray}
where we have taken $(d^{2}/dt_{1}^{2})(d^{2}/dt_{2}^{2})G(t_{1},t_{2})=(d^{2}/dt_{1}^{2})\delta_{\lambda}(t_{1}-t_{2})$. Since the dissipation kernel does not depend on the temperature, we are left with two cutoff scales $\Lambda$ and $\lambda$. As before, we assume that the cutoff scale of the gravitons are larger than that of the quantum particles, that is, $\Lambda\gg\lambda$. Under this condition, we have the expansion
\begin{eqnarray}
    d_{1}(t)=\frac{512\pi^{4}\alpha^{2}}{3m}\left[\left(-\frac{\lambda^{3}}{3}\right)\left(\frac{1}{\pi}\Lambda\,t-\frac{1}{2}\left(\theta(t)-\theta(-t)\right)\right)+\cdots\right]
\end{eqnarray}
Similarly, we also have
\begin{eqnarray}
    d_{2}(t)&=&\frac{512\pi^{4}\alpha^{2}}{3m^{2}}\left[\left(-\frac{\lambda^{3}t}{3}\right)\left(\frac{1}{\pi}\Lambda\,t-\frac{1}{2}\left(\theta(t)-\theta(-t)\right)\right)+\cdots\right]\nonumber\\
    d_{3}(t)&=&\frac{512\pi^{4}\alpha^{2}}{3m^{3}}\left[\frac{\Lambda}{\pi}\left(\frac{1}{9}\lambda^{3}t^{3}+8\lambda\,t\right)-\lambda\left(\frac{1}{12}\lambda^{2}t^{2}+4(\theta(t)-\theta(-t))\right)+\dots\right]\nonumber\\
    d_{4}(t)&=&\frac{2048\pi^{4}\alpha^{2}}{15m^{4}}\left[\Lambda\,t-\frac{\pi}{2}(\theta(t)-\theta(-t))+\cdots\right]
\end{eqnarray}

\subsection{The master equation}
The coefficient functions of the master equation in Eq.~(\ref{mastereqn1}) are listed in Eq.~(\ref{mastercof}). The coefficients related to $n_{1}(t)$ to $n_{4}(t)$ are temperature dependent. Let us examine   their low and high temperature limits. First, we consider the low temperature expansion  with $\Lambda\gg\lambda\gg 1/t\gg 1/\beta$.
\begin{eqnarray}
    \frac{1}{m_{\rm ren}}
    &=&\frac{1}{m}(1+\dot{n}_{3}(t))\nonumber\\
    &=&\frac{1}{m}+\frac{256\pi^{4}\alpha^{2}}{3m^{3}}\Bigg\{\bigg[\lambda^{2}\left(\frac{1}{4}\lambda^{2}t^{2}-\frac{4}{9\pi}\lambda\,t+7\right)+\frac{\lambda^{2}}{\pi}\left(\frac{2}{3}\lambda\,t\right)({\rm ln}\Lambda-{\rm ln}\lambda)\bigg]\nonumber\\
    &&\hskip 90pt +\frac{1}{\beta^{2}}\left(\frac{2\pi}{3}\right)(\lambda\,t+\pi)+\frac{1}{\beta^{4}}\left(\frac{2\pi^{4}t^{2}}{15}\right)+\cdots\Bigg\}
\end{eqnarray}
\begin{eqnarray}
    N_{1}(t)&=&\dot{n}_{1}(t)+\frac{3}{m}n_{2}(t)\nonumber\\
    &=&\frac{256\pi^{4}\alpha^{2}}{m^{2}}\Bigg\{\bigg[\frac{8}{t^{3}}({\rm ln}(\Lambda\,t)+\gamma)-\frac{\lambda}{\pi t^{2}}\left(\frac{1}{3}\lambda^{2}t^{2}+16\right)({\rm ln}\Lambda-{\rm ln}\lambda)\nonumber\\
    &&\hskip 60pt +\frac{\lambda}{t^{2}}\left(-\frac{1}{12}\lambda^{3}t^{3}+\frac{2}{9\pi}\lambda^{2}t^{2}+\frac{35}{3\pi t^{2}}\right)-\frac{2}{\pi t^{3}}\sin(\lambda\,t)\bigg]\nonumber\\
    &&\hskip 60pt+\frac{1}{\beta^{2}}\left(\frac{\pi}{3\,t}\right)\left(-\lambda\,t+2\sin(\lambda\,t)\right)-\frac{1}{\beta^{4}}\left(\frac{2\pi^{3}t}{15}\sin(\lambda\,t)\right)\nonumber\\
    &&\hskip 60pt+\frac{1}{\beta^{6}}\left(\frac{4\pi^{5}t^{3}}{189}\right)\left(-\frac{4\pi}{15}+\sin(\lambda\,t)\right)+\cdots\Bigg\}
\end{eqnarray}
\begin{eqnarray}
    N_{2}(t)&=&\dot{n}_{2}(t)\nonumber\\
    &=&\frac{256\pi^{4}\alpha^{2}}{3m}\Bigg\{\bigg[-\frac{1}{4}\lambda^{4}-\frac{2}{\pi t^{4}}\left(\lambda\,t\cos(\lambda\,t)-3\sin(\lambda\,t)\right)\bigg]\nonumber\\
    &&\hskip 70pt+\frac{1}{\beta^{2}}\left(\frac{2\pi}{3t^{2}}\right)\left(\lambda\,t\cos(\lambda\,t)-\sin(\lambda\,t)\right)\nonumber\\
    &&\hskip 70pt-\frac{1}{\beta^{4}}\left(\frac{2\pi^{3}}{15}\right)\left(\pi+\lambda\,t\cos(\lambda\,t)+\sin(\lambda\,t)\right)\nonumber\\
    &&\hskip 70pt+\frac{1}{\beta^{6}}\left(\frac{4\pi^{5}t^{2}}{63}\right)\left(\frac{1}{3}\lambda\,t\cos(\lambda\,t)+\sin(\lambda\,t)\right)+\cdots\Bigg\}
\end{eqnarray}
\begin{eqnarray}
    N_{3}(t)&=&\frac{1}{m}n_{2}(t)\nonumber\\
    &=&\frac{256\pi^{4}\alpha^{2}}{3m^{2}}\Bigg\{\bigg[-\frac{2}{3\pi}\lambda^{3}\left({\rm ln}\Lambda-{\rm ln}\lambda\right)+\lambda^{3}\left(-\frac{1}{4}\lambda\,t+\frac{4}{9\pi}\right)-\frac{2}{\pi t^{3}}\sin(\lambda\,t)\bigg]\nonumber\\
    &&\hskip 70pt+\frac{1}{\beta^{2}}\left(\frac{2\pi}{3t}\right)\left(-\lambda\,t+\sin(\lambda\,t)\right)-\frac{1}{\beta^{4}}\left(\frac{2\pi^{3}t}{15}\right)\left(\pi+\sin(\lambda\,t)\right)\nonumber\\
    &&\hskip 70pt+\frac{1}{\beta^{6}}\left(\frac{4\pi^{5}t^{3}}{189}\right)\sin(\lambda\,t)+\cdots\Bigg\}
\end{eqnarray}
\begin{eqnarray}
    N_{4}(t)&=&\dot{n}_{4}(t)\nonumber\\
    &=&-\frac{512\pi^{4}\alpha^{2}}{15m^{4}}\left[\frac{1}{t}+\frac{\pi^{2}t}{3\beta^{2}}-\frac{\pi^{4}t^{3}}{45\beta^{4}}+\frac{2\pi^{6}t^{5}}{945\beta^{6}}+\cdots\right]
\end{eqnarray}

For the high temperature expansion with $1/\beta\gg\Lambda\gg\lambda\gg 1/t$, we have
\begin{eqnarray}
    \frac{1}{m_{\rm ren}}&=&\frac{1}{m}+\frac{256\pi^{4}\alpha^{2}}{3m^{3}}\Bigg\{\frac{\lambda}{\beta}\left(\frac{2}{3}\lambda^{2}t^{2}+16\right)\nonumber\\
    &&\hskip 45pt +\beta\bigg[\frac{\Lambda\lambda^{3}t}{9\pi}+\frac{\lambda^{3}}{9}\left(\frac{3}{10}\lambda^{2}t^{2}+10\right)-\frac{2\Lambda}{\pi t^{2}}\cos(\Lambda\,t)\left(\lambda\,t\cos(\lambda\,t)+\frac{1}{3}\sin(\lambda\,t)\right)\nonumber\\
    &&\hskip 80pt+\frac{2}{\pi t^{3}}\sin(\Lambda\,t)\left(\lambda^{2}t^{2}\sin(\lambda\,t)+\frac{2}{3}\lambda\,t\cos(\lambda\,t)+\frac{2}{3}\sin(\lambda\,t)\right)\bigg]+\cdots\Bigg\}\nonumber\\
\end{eqnarray}
\begin{eqnarray}
    N_{1}(t)&=&\frac{256\pi^{4}\alpha^{2}}{m^{2}}\Bigg\{\frac{1}{\beta}\left(\frac{2}{t^{2}}\right)\left(-\frac{1}{9}\lambda^{3}t^{3}+2\pi\right)\nonumber\\
    &&\hskip 45pt +\beta\bigg[\frac{\Lambda}{3\pi t^{3}}\left(-\frac{1}{6}\lambda^{3}t^{3}-8\lambda\,t+4\pi\right)-\frac{\lambda^{3}}{t}\left(\frac{1}{90}\lambda^{2}t^{2}\right)\nonumber\\
    &&\hskip 80pt+\frac{2\Lambda}{3\pi t^{3}}\cos(\Lambda\,t)\left(\pi-3\lambda\,t\cos(\lambda\,t)-\sin(\lambda\,t)\right)\bigg]\nonumber\\
    &&\hskip 80pt+\frac{1}{\pi t^{4}}\sin(\Lambda\,t)\left(\frac{7}{3}\lambda^{2}t^{2}\sin(\lambda\,t)+4\lambda\,t\cos(\lambda\,t)+\frac{4}{3}\sin(\lambda\,t)-2\pi\right)\nonumber\\
    &&\hskip 100pt+\cdots\Bigg\}
\end{eqnarray}
\begin{eqnarray}
    N_{2}(t)&=&\frac{256\pi^{4}\alpha^{2}}{3m}\Bigg\{-\frac{1}{\beta}\left(\frac{2}{3}\lambda^{3}\right)+\beta\bigg[-\frac{1}{30}\lambda^{5}\nonumber\\
    &&\hskip 5pt+\frac{\Lambda}{3\pi t^{4}}\cos(\Lambda\,t)\left(\lambda^{2}t^{2}\sin(\lambda\,t)+2\lambda\,t\cos(\lambda\,t)-2\sin(\lambda\,t)\right)\nonumber\\
    &&\hskip 5pt+\frac{1}{3\pi t^{5}}\sin(\Lambda\,t)\left(\lambda^{3}t^{3}\cos(\lambda\,t)-4\lambda^{2}t^{2}\sin(\lambda\,t)-8\lambda\,t\cos(\lambda\,t)+8\sin(\lambda\,t)\right)\bigg]+\cdots\Bigg\}\nonumber\\
\end{eqnarray}
\begin{eqnarray}
    N_{3}(t)&=&\frac{256\pi^{4}\alpha^{2}}{3m^{2}}\Bigg\{-\frac{1}{\beta}\left(\frac{2}{3}\lambda^{3}t\right)+\beta\bigg[-\frac{1}{9\pi}\Lambda\lambda^{3}-\frac{1}{30}\lambda^{5}t\nonumber\\
    &&\hskip 70pt+\frac{1}{3\pi t^{4}}\sin(\Lambda\,t)\left(\lambda^{2}t^{2}\sin(\lambda\,t)+2\lambda\,t\cos(\lambda\,t)-2\sin(\lambda\,t)\right)\bigg]+\cdots\Bigg\}\nonumber\\
\end{eqnarray}
\begin{eqnarray}
    N_{4}(t)=-\frac{256\pi^{4}\alpha^{2}}{15m^{4}}\bigg[\frac{2\pi}{\beta}-\beta\left(\frac{1}{3t^{2}}\right)\left(\Lambda\,t\cos(\Lambda\,t)-\sin(\Lambda\,t)\right)+\cdots\bigg]
\end{eqnarray}

The coefficient functions $D_{1}$ to $D_{4}$ related to the dissipation kernel are independent of temperature but still we can get their expressions under the limit $\Lambda\gg\lambda\gg 1/t$ 
\begin{eqnarray}
    D_{1}(t)=\dot{d}_{1}(t)=\frac{512\pi^{5}\alpha^{2}}{3m}\left[-\left(\frac{\lambda^{3}}{3\pi}\right)\left(\frac{\Lambda}{\pi}-\delta(t)\right)+\cdots\right]
\end{eqnarray}
\begin{eqnarray}
    D_{2}(t)=\dot{d}_{2}(t)-\frac{2}{m}d_{1}(t)=\frac{512\pi^{5}\alpha^{2}}{3m^{2}}\left[\left(\frac{\lambda^{3}}{3\pi}\right)\left(\frac{1}{2}-(\theta(t)-\theta(-t))\right)+\cdots\right]
\end{eqnarray}
\begin{eqnarray}
    D_{3}(t)=\dot{d}_{3}(t)+\frac{1}{m}d_{2}(t)=\frac{512\pi^{5}\alpha^{2}}{3m^{3}}\left[\left(\frac{8\lambda}{\pi}\right)\left(\frac{\Lambda}{\pi}-\delta(t)\right)+\cdots\right]
\end{eqnarray}
\begin{eqnarray}
    D_{4}(t)=\dot{d}_{4}(t)=\frac{2048\pi^{5}\alpha^{2}}{15m^{4}}\left[\left(\frac{\Lambda}{\pi}-\delta(t)\right)+\cdots\right]
\end{eqnarray}

\end{document}